\pdfoutput=1
\documentclass[%
 reprint,
 amsmath,amssymb,
 aps,
]{revtex4-1}

\usepackage{graphicx}
\usepackage{dcolumn}
\usepackage{bm}
\usepackage{color}

\newcommand{\RN}[1]{%
  \textup{\uppercase\expandafter{\romannumeral#1}}%
}

\newcommand{\km}[1]{\color{magenta}}

\begin{document}
\preprint{APS/123-QED}

\title{Low-luminosity gamma-ray bursts as the sources of ultrahigh-energy cosmic ray nuclei}

\author{B. Theodore Zhang$^{1,2,3}$}
\author{Kohta Murase$^{3,4,5,6}$}
\author{Shigeo S. Kimura$^{3, 4, 5}$}
\author{Shunsaku Horiuchi$^{7,8}$}
\author{Peter M{\'e}sz{\'a}ros$^{3, 4, 5}$}

\affiliation{$^1$Department of Astronomy, School of Physics, Peking University, Beijing 100871, China }
\affiliation{$^2$Kavli Institute for Astronomy and Astrophysics, Peking University, Beijing 100871, China}
\affiliation{$^3$Department of Physics, The Pennsylvania State University, University Park, Pennsylvania 16802, USA}
\affiliation{$^4$Department of Astronomy \& Astrophysics, The Pennsylvania State University, University Park, Pennsylvania 16802, USA}
\affiliation{$^5$Center for Particle and Gravitational Astrophysics, The Pennsylvania State University, University Park, Pennsylvania 16802, USA}
\affiliation{$^6$Center for Gravitational Physics, Yukawa Institute for Theoretical Physics, Kyoto, Kyoto 606-8502 Japan}
\affiliation{$^7$Department of Physics, Virginia Tech, Blacksburg, VA 24060}
\affiliation{$^8$Center for Neutrino Physics, Department of Physics, Virginia Tech, Blacksburg, VA 24060}

\date{\today}
             
\begin{abstract}
Recent results from the Pierre Auger Collaboration have shown that the composition of ultrahigh-energy cosmic rays (UHECRs) becomes gradually heavier with increasing energy. Although gamma-ray bursts (GRBs) have been promising sources of UHECRs, it is still unclear whether they can account for the Auger results because of their unknown nuclear composition of ejected UHECRs. In this work, we revisit the possibility that low-luminosity GRBs (LL GRBs) act as the sources of UHECR nuclei, and give new predictions based on the intrajet nuclear composition models considering progenitor dependencies. We find that the nuclear component in the jet can be divided into two groups according to the mass fraction of silicon nuclei, Si-free and Si-rich. Motivated by the connection between LL GRBs and transrelativistic supernovae, we also consider the hypernova ejecta composition. Then, we discuss the survivability of UHECR nuclei in the jet base and internal shocks of the jets, and show that it is easier for nuclei to survive for typical LL GRBs. Finally, we numerically propagate UHECR nuclei ejected from LL GRBs with different composition models and compare the resulting spectra and composition to Auger data. Our results show that both the Si-rich progenitor and hypernova ejecta models match the Auger data well, while the Si-free progenitor models have more difficulty in fitting the spectrum. We argue that our model is consistent with the newly reported cross correlation between the UHECRs and starburst galaxies, since both LL GRBs and hypernovae are expected to be tracers of the star-formation activity. LL GRBs have also been suggested as the dominant origin of IceCube neutrinos in the PeV range, and the LL GRB origin of UHECRs can be critically tested by near-future multimessenger observations. 
\end{abstract}

\pacs{Valid PACS appear here}
\maketitle

\section{\label{sec:one}Introduction}
The origin of ultrahigh-energy cosmic rays (UHECRs) is a long standing unresolved problem in astroparticle physics~\cite{Linsley:1963km, Hillas:1985is, Nagano:2000ve, Kotera:2011cp}. It has been widely believed that these particles have an extragalactic origin. This argument is supported by the observed flux suppression at the highest energy which is consistent with the prediction of Greisen-Zatsepin-Kuzmin (GZK) cutoff~\cite{Abraham:2008ru, Abbasi:2007sv} and the observed large-scale anisotropy described by a dipole with cosmic rays have energy above $8 \times 10^{18}~\rm eV$~\cite{Aab:2017tyv}. 
The composition of UHECRs is the key to identifying the sources of these rare particles. The primary mass of UHECRs can be inferred from the depth of shower maximum, $X_{\rm max}$~\cite{Aab:2014aea, Aab:2014kda}. Recent data sets from fluorescence telescopes and water-Cherenkov detectors of the Auger observatory have shown that the primary mass of UHECRs is dominated by light composition at energy of around $2 \times 10^{18}~\rm eV$, and then becomes gradually heavier with increasing energy up to $4 \times 10^{19}~\rm eV$, and may stop increasing beyond this energy~\cite{Aab:2017njo}. The interpretation of the UHECR composition is controversial, and the Telescope Array (TA) collaboration has claimed that their data prefer light composition dominated scenario~\cite{Fukushima:2015bza, Abbasi:2018nun}. 
It has been shown that the $\left\langle X_{\rm max}\right\rangle$ data from the TA and Auger are in good agreement within systematic and statistical uncertainties~\cite{Abbasi:2015xga, TheTelescopeArray:2018dje}. It is reasonable to assume that UHECRs are dominated by heavy nuclei considering the larger detection area and lower sampling bias of Auger, which have been supported independently by the non-observation of cosmogenic neutrinos~\cite{Aab:2015kma, Aartsen:2016ngq}.

There are three main implications about the sources if UHECRs are dominated by heavy nuclei. First, the source material should be rich in heavy nuclei. According to the standard stellar evolution model, compact stars (white dwarfs and the atmosphere of neutron stars) or the inner cores of massive stars are mainly composed of heavy nuclei. Such objects are related to a number of well-motivated ``intermediate or heavy composition'' sources, such as
gamma-ray bursts~\cite{Murase:2008mr, Wang:2007xj,Horiuchi:2012by, Metzger:2011xs, Globus:2014fka, Biehl:2017zlw}, transrelativistic supernovae or engine-driven supernovae~\cite{Wang:2007ya, Murase:2008mr, Chakraborti:2010ha, Liu:2011tv}, new-born pulsars~\cite{Fang:2012rx, Fang:2013vla, Kotera:2015pya}, and tidal disruption events~\cite{AlvesBatista:2017shr, Zhang:2017hom, Biehl:2017hnb, Guepin:2017abw}. Heavy nuclei are abundant in Galactic TeV--PeV cosmic rays, and re-acceleration of them in radio galaxies are proposed as an alternative possibility~\cite{Caprioli:2015zka, Kimura:2017ubz}. Second, the sources environment should allow nuclei to survive (e.g.,~\cite{Murase:2008mr, Wang:2007xj, Murase:2010gj}). This requirement limits the ability of luminous objects to act as the sources of UHECR nuclei since most of them may be destroyed due to the interaction with background photons or protons. Finally, there should be enough sources located in the very nearby region within $\sim10-100\rm~Mpc$ around our Galaxy, otherwise many of the UHECR nuclei would be photodisintegrated into free nucleons before reaching Earth.

GRBs are the brightest and probably the most powerful high energy astrophysical phenomenon in the universe~\cite{Meszaros:2006rc, Woosley:2006fn, Kumar:2014upa}. It seems that nearly all the GRBs are accompanied by energetic broadline Type \RN{1}c supernovae (or hypernovae for objects with a kinetic energy of $\gtrsim{10}^{52}$~erg)~\cite{Woosley:2006fn}. The GRB-SN connection reveals the fact that both of them are related to the deaths of massive stars. The observed GRBs can be divided into two groups: high-luminosity GRBs (HL GRBs) which have typical isotropic radiation luminosity in the range from $10^{51} \rm~erg~s^{-1}$ to $10^{53} \rm~erg~s^{-1}$ and low-luminosity GRBs (LL GRBs) with luminosity less than $\sim10^{49} \rm~erg~s^{-1}$~\cite{Liang:2006ci, Virgili:2008gp, Sun:2015bda}. It is usually believed that LL GRBs share similar characteristics with the classical HL GRBs except that their relativistic outflows have lower Lorentz factors (that could be caused by nearly choked jets or large baryon contamination).  
The conventional HL GRBs have been suggested as the sources of UHECRs considering that the charged particles (mainly proton dominated) can be accelerated to ultrahigh energies $\sim~10^{20}\rm~eV$ and satisfy the required energy budget of UHECRs \cite{Waxman:1995vg,Vietri:1995hs}. 
Murase et al.~\citep{Murase:2006mm,Murase:2008mr} first proposed that LL GRBs can be the sources of UHECRs and they are preferred to be the sources of UHECR nuclei compared to HL GRBs. The advantages for LL GRBs are that they are likely to have a much higher event rate in the local universe~\cite{Soderberg:2006vh, Liang:2006ci, Sun:2015bda} and heavy nuclei much easily survive inside the sources due to their lower radiation luminosity~\cite{Murase:2008mr, Horiuchi:2012by}. 

However, the previous works that investigated the contribution from LL GRBs did not attempt to fit the UHECR spectrum and composition, so it is still unclear whether UHECR nuclei originating from LL GRBs can quantitatively account for both the spectrum and composition of UHECRs measured by Auger. One of the uncertainties comes from the unknown mass fraction of UHECR nuclei ejected from LL GRBs. It has been shown that the jets of LL GRBs can readily contain a significant fraction of nuclei~\cite{Horiuchi:2012by,Shibata:2015mva}. In this work, we aim to give new predictions based on nuclear composition models inside the jet, considering progenitor dependencies, and examine whether they are compatible with Auger results. 

In the multi-messenger astronomy era, both the neutrinos and photons are helpful for identifying the sources of UHECRs~\cite{Kotera:2011cp, Meszaros:2017nhc}. High-energy astrophysical neutrinos in the TeV-PeV energy range have been detected by IceCube~\cite{Aartsen:2013bka, Aartsen:2013jdh, Aartsen:2014gkd, Aartsen:2015rwa}. The stacking analysis has challenged the possibility that conventional HL GRBs make a significant contribution to the diffuse neutrino flux observed by IceCube~\cite{Aartsen:2014aqy,Aartsen:2016qcr}. 
However, LL GRBs, which are dim but numerous than the conventional HL GRBs, remain viable to be the dominant sources of IceCube neutrinos~\cite{Murase:2015xka}. Indeed, one of the predictions in Ref.~\cite{Murase:2006mm} is consistent with the IceCube data above $\sim0.1$~PeV~\cite{Murase:2015xka}, and their choked jet population can significantly contribute to the diffuse neutrino flux even in the $10-100$~TeV range~\cite{Murase:2013ffa, Senno:2015tsn}. 

This paper is organized as follows. In Section~\ref{sec:two}, we  estimate the nuclear mass fraction in the jet of LL GRBs. In Section~\ref{sec:three}, we briefly discuss the acceleration and survival of UHECR nuclei in GRBs, and consider the escape process to get the injection spectrum of UHECR nuclei. In Section~\ref{sec:four}, we numerically propagate UHECR nuclei ejected from LL GRBs using publicly available Monte Carlo code CRPropa 3~\cite{Batista:2016yrx} and compare our results to the Auger data. In Section~\ref{sec:five}, we discuss the implications of the LL GRBs models to IceCube neutrinos.
Finally, we give the conclusion and discussion in Section~\ref{sec:six}.

Throughout of the paper, we adopt the cgs unit, and have notations $Q_x \equiv Q /10^x$. The cosmological parameters we use are $H_0 = 67.3 \ \rm km \ s^{-1} \ Mpc^{-1}$, $\Omega_m = 0.315$, $\Omega_\Lambda = 0.685$ \cite{Agashe:2014kda}.
 
\section{\label{sec:two} Intrajet nuclear composition}
It is widely accepted that the heavy nuclei up to iron are synthesized in the inner core of massive stars~\cite{Woosley:2002zz}. The long GRBs can be explained in the collapsar scenario~\cite{Woosley:1993wj, MacFadyen:1998vz}, where the inner iron core eventually collapses into a black hole and the surrounding stellar material initiates fall back until it can be supported by the centrifugal force. A relativistic outflow along the polar direction can be driven after the formation of an accretion disk around the central compact remnant (either a black hole or magnetar), and it breaks out from the stellar surface to appear as an GRB. However, it is still unclear how to convert a fraction of the gravitational or rotational energy into the relativistic outflow. One of the possible mechanisms is related to the the hyper accretion disk where neutrino annihilation can deposit enormous energy in the vicinity of the remnant~\cite{MacFadyen:1998vz, Popham1999a, Narayan:2001qi, Matteo:2002ck, Kohri:2002kz, Lei:2012df}. An alternative scenario involves MHD processes involving the amplification of pre-existing magnetic fields. In this case, for the collapsar scenario, the energy of the relativistic outflow can be extracted from the surface of the accretion disk ~\cite{Blandford:1982di} or from the rotating black hole via the Blandford-Znajek mechanism~\cite{Blandford:1977ds}. Note that the central engine may not necessarily be a black hole but could alternatively be a fast rotating and strongly magnetized protoneutron star~\cite{Usov:1992zd}.  


In this work, we assume that the initial loading at the jet base is the dominant process for the nuclear composition in the relativistic outflow. It is directly related to the presupernova models, which usually have an ``onion-skin'' structure at the final evolution stage with different layers composed of different nuclei species. The initial loading process can be simplified to the ``one-time'' injection scenario. In this scenario, the falling stellar nuclei form an accretion disk at first, and then a fraction of them are picked up by the relativistic outflow. The detailed physical processes of the initial loading are related to the unknown diffusive processes or the geometry of open magnetic fields~\cite{Levinson:2003je, Metzger:2011xs}. The jet nuclei can also be altered by the entrainment during the jet propagation inside the progenitors~\cite{Horiuchi:2012by} or explosive nucleosynthesis if the entropy is lower~\cite{Metzger:2011xs}.
 
\subsection{\label{sec:two_1}GRB progenitors}
The association of long GRBs with Type \RN{1}c-BL SN reveals the fact that GRBs are the outcome of the deaths of massive stars. Wolf-Rayet stars (WRs), which have expelled their hydrogen and/or helium envelop during their evolution from main-sequence stage, are believed to be the progenitors of Type \RN{1}c SN. The production of GRBs is also possible if WRs have enough angular momentum in their inner core at the onset of core collapse. However, it seems difficult for the progenitors to maintain enough angular momentum during the evolution even though they rotate very fast initially due to the wind mass loss process or the effect of magnetic torques~\cite{Woosley:2005gy}. One of the possible solutions is to consider the chemically homogeneous evolution (CHE) scenario~\cite{Woosley:2005gy, Yoon:2005tv, Yoon:2006fr, Dessart:2017jib}, where the products of nuclear reaction are completely mixed during the main sequence stage in rapidly rotating stars~\cite{Heger:1999ax}. Also, stars undergoing CHE process will not go through the red super giants (RSGs) phase and evolves to WRs directly~\cite{Woosley:2005gy}. 

In this work, we choose five typical presupernova models calculated by Ref.~\cite{Woosley:2005gy}, which are evolved from stars with different initial mass, metallicity, and rotation speed. The progenitors of GRBs can be bare helium cores as the result of binary interaction or rapidly rotating single stars. We summarize the basic properties of the five presupernova models in Table~\ref{tab:table1}. 

\begin{table*}
\caption{\label{tab:table1}Jet nuclei composition models}
\begin{ruledtabular}
\begin{tabular}{lccccccccccccc}
 \multicolumn{1}{c}{Models\footnotemark[1]}&\multicolumn{1}{c}{$M_{\rm init}$\footnotemark[2]}&\multicolumn{1}{c}{$M_{\rm final}$\footnotemark[3]}&\multicolumn{1}{c}{${\mathcal J}_{\rm core}$\footnotemark[4]}&\multicolumn{1}{c}{$r_c$\footnotemark[5]}&\multicolumn{1}{c}{$M_c$\footnotemark[6]}&\multicolumn{6}{c}{Jet nuclei composition\footnotemark[7]}\\
    &$M_\odot$&$M_\odot$&$10^{47} \ \rm erg \ s$&$10^9~\rm cm$&$M_\odot$&C &O & Ne & Mg & Si & S  & Fe\\ \hline
 
Si-F 1 (HE16F) &16 &$14.80$ &$114$   &$1.9$ &$4.1$ &$ 0.018$&$0.698$&$0.243$&$0.036$& & \\
Si-F 2 (16TI)   &16 &$13.95$ & $87$ &$2.0$  &$3.3$ &$0.022$&$0.695$&$0.247$&$0.034$ & & \\
Si-R 1 (12TJ)  &12 &$11.54$ & $150$  &$0.5$  &$2.5$ & &$0.603$& & &$0.351$&$0.046$ \\
Si-R 2 (16TJ)  &16 &$15.21$ & $178$  &$0.6$  &$2.5$ & &$0.511$& & &$0.364$&$0.108$\\
Si-R 3 (35OC) &35 &$28.07$ & $230$  &$1.2$  &$3.9$ & &$0.157$& & & $0.421$&$0.303$ \\
Hypernova &-- &-- & --&--&-- & $0.006$&$0.710$&$0.036$ & $0.034$& $0.083$&$0.041$&$0.090$
\end{tabular}
\end{ruledtabular}
\footnotetext[1]{Presupernova models calculated in Ref.~\cite{Woosley:2005gy}.}
\footnotetext[2]{The initial mass of GRBs progenitors.}
\footnotetext[3]{The final mass of GRBs progenitors at the onset of core collapse.}
\footnotetext[4]{The angular momentum of the iron core at core collapse.}
\footnotetext[5]{Critical radius in the progenitors where accreting material starts to form the accretion disk.}
\footnotetext[6]{Enclosed mass within the critical radius $r_c$.}
\footnotetext[7]{Jet nuclear composition. The blank space means that nuclei have mass fraction less than 0.001. The last row represents the hypernova ejecta composition.}
\end{table*}

\begin{figure}
\includegraphics[width=\linewidth]{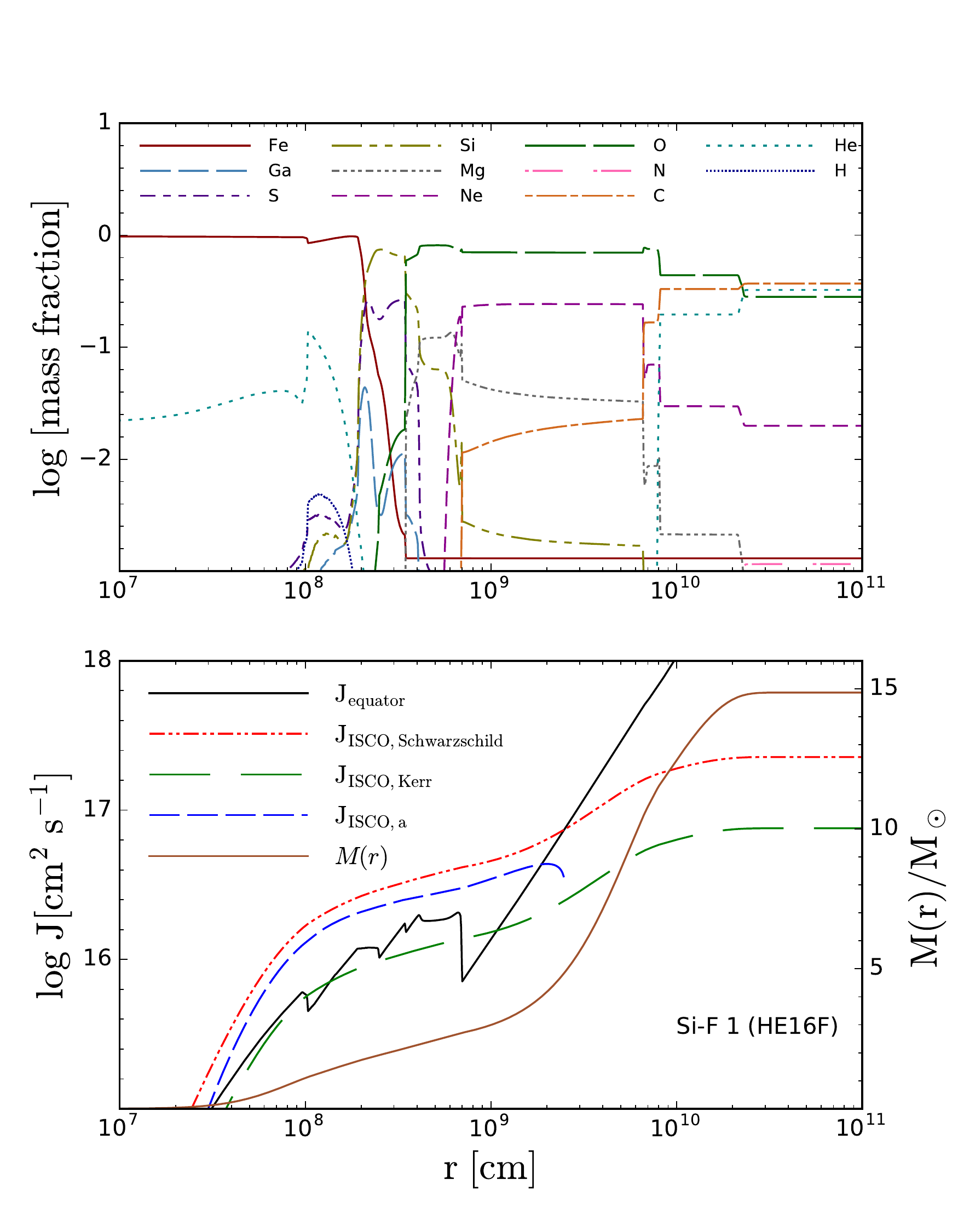}
\caption{The distribution of the nuclear mass fraction (upper panel) and the specific angular momentum $J_{\rm equator}$ (lower panel) for presupernova model Si-F 1 (HE16F) at the onset of core collapse. The red, green, and blue lines are the estimated specific angular momentum at ISCO as a function of black hole mass, see the definition in the main text. We also show the distribution of enclosed mass as a function of radius $r$ in a brown line. The Si-F 1 (HE16F) model is evolved from a bare helium core with an initial mass $16 M_\odot$ as the outcome of the stellar merger. The final mass at core collapse is $14.80 M_\odot$ considering a fraction of $0.03 \dot{M}$ of the standard WR mass-loss rate and the Fe core has an angular momentum ${\mathcal J}_{\rm Fe~core} = 114 \times 10^{47}\rm~erg~s$ at core collapse after considering the effect of magnetic torques~\cite{Woosley:2005gy}. \label{fig:HE16F}}
\end{figure}

\begin{figure}
\includegraphics[width=\linewidth]{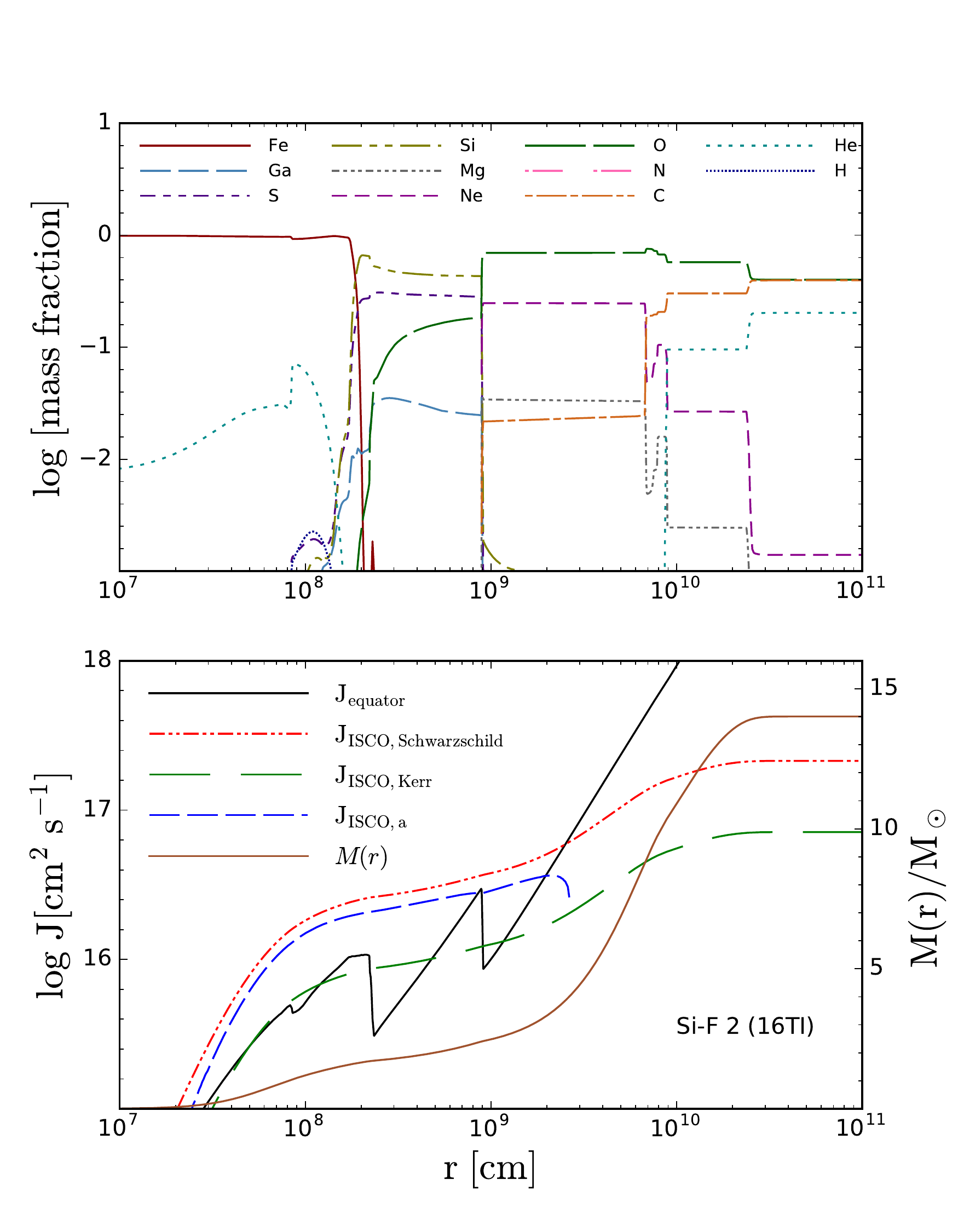}
\caption{Same as Fig.~\ref{fig:HE16F}, but for presupernova model Si-F 2 (16TI). The Si-F 2 (16TI) model are evolved from a rapidly rotating star with initial mass $16 M_\odot$ and $1\%$ solar metallicity. The final mass is $13.95 M_\odot$ considering a fraction of $0.3 \dot{M}$ of the standard WR mass-loss rate and the Fe core has an angular momentum at core collapse ${\mathcal J}_{\rm Fe~core} = 86.7 \times 10^{47}\rm~erg~s$ considering the effect of magnetic torques~\cite{Woosley:2005gy}. \label{fig:16TI}}
\end{figure}

\begin{figure}
\includegraphics[width=\linewidth]{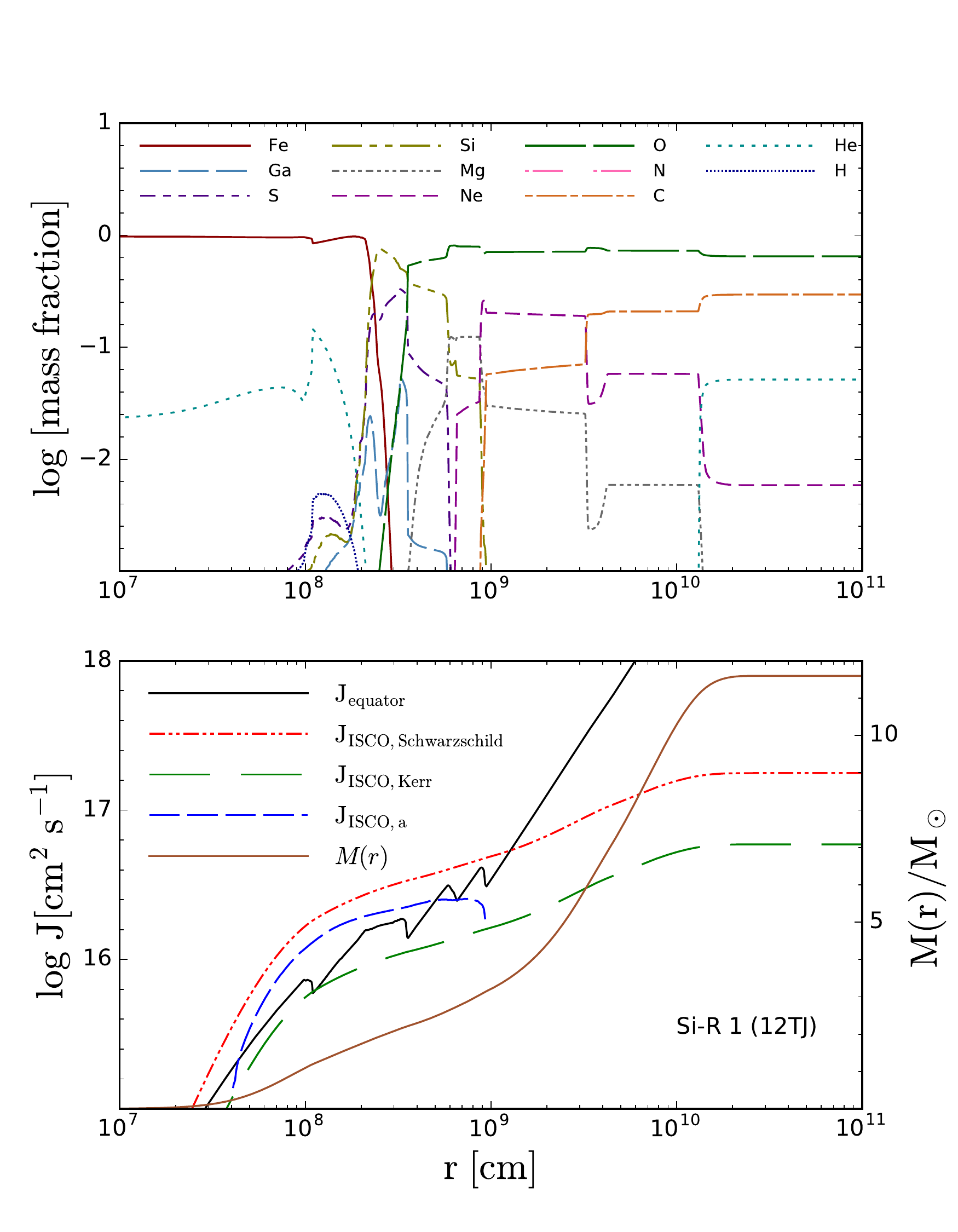}
\caption{Same as Fig.~\ref{fig:HE16F}, but for presupernova model Si-R 1 (12TJ). The Si-R 1 (12TJ) model are evolved from a rapidly rotating star with initial mass $12 M_\odot$ and $1\%$ solar metallicity. The final mass is $11.54 M_\odot$ considering a fraction of $0.1 \dot{M}$ of the standard WR mass-loss rate and the Fe core has an angular momentum at core collapse ${\mathcal J}_{\rm Fe~core} = 150 \times 10^{47}\rm~erg~s$ considering the effect of magnetic torques~\cite{Woosley:2005gy}. \label{fig:12TJ}}
\end{figure}

\begin{figure}
\includegraphics[width=\linewidth]{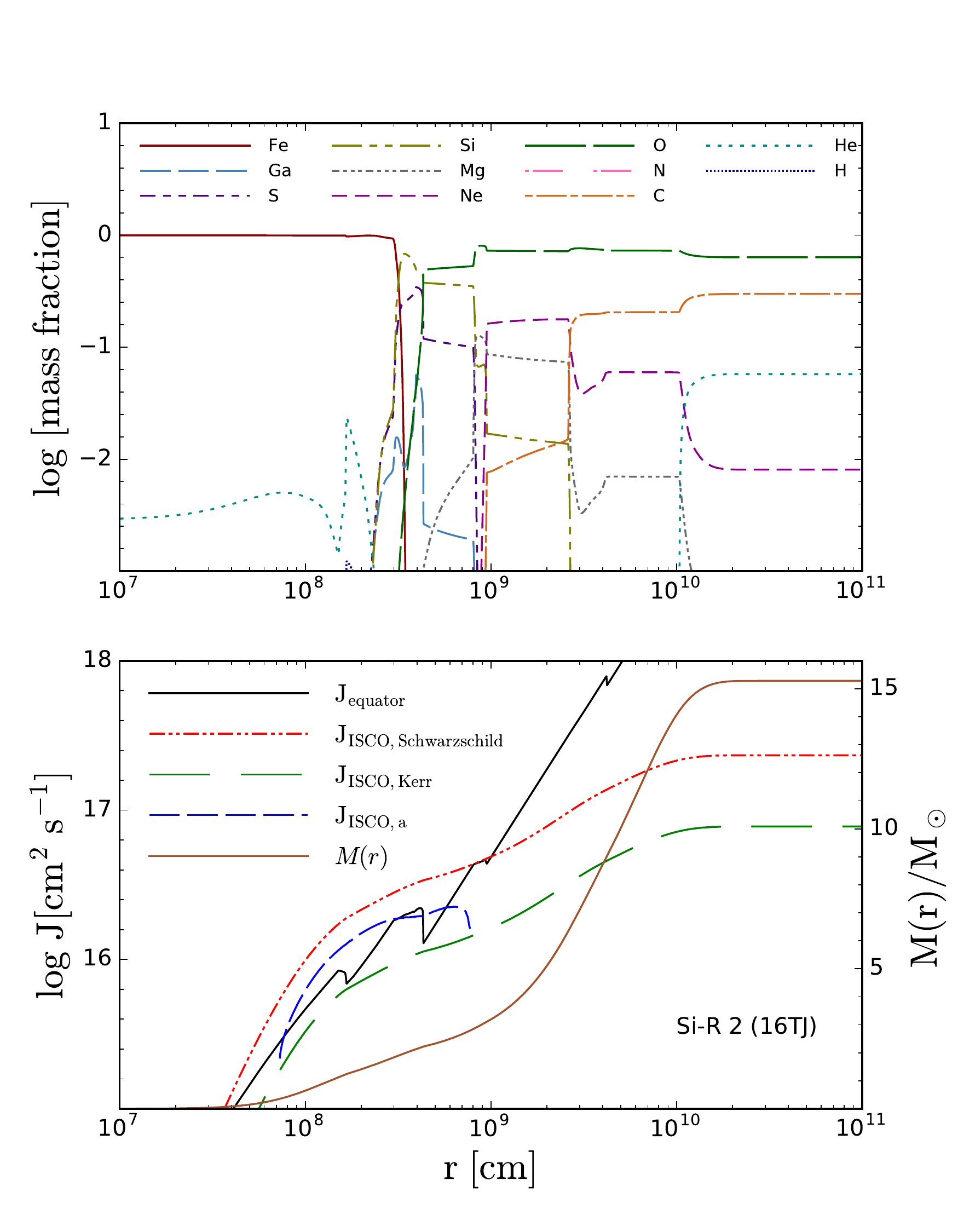}
\caption{Same as Fig.~\ref{fig:HE16F}, but for presupernova model Si-R 2 (16TJ). The Si-R 2 (16TJ) model are evolved from a rapidly rotating star with initial mass $16 M_\odot$ and $1\%$ solar metallicity. The final mass is $15.21 M_\odot$ considering a fraction of $0.1 \dot{M}$ of the standard WR mass-loss rate and the Fe core has an angular momentum at core collapse ${\mathcal J}_{\rm Fe~core} = 178 \times 10^{47}\rm~erg~s$ considering the effect of magnetic torques~\cite{Woosley:2005gy}. \label{fig:16TJ}}
\end{figure}

\begin{figure}
\includegraphics[width=\linewidth]{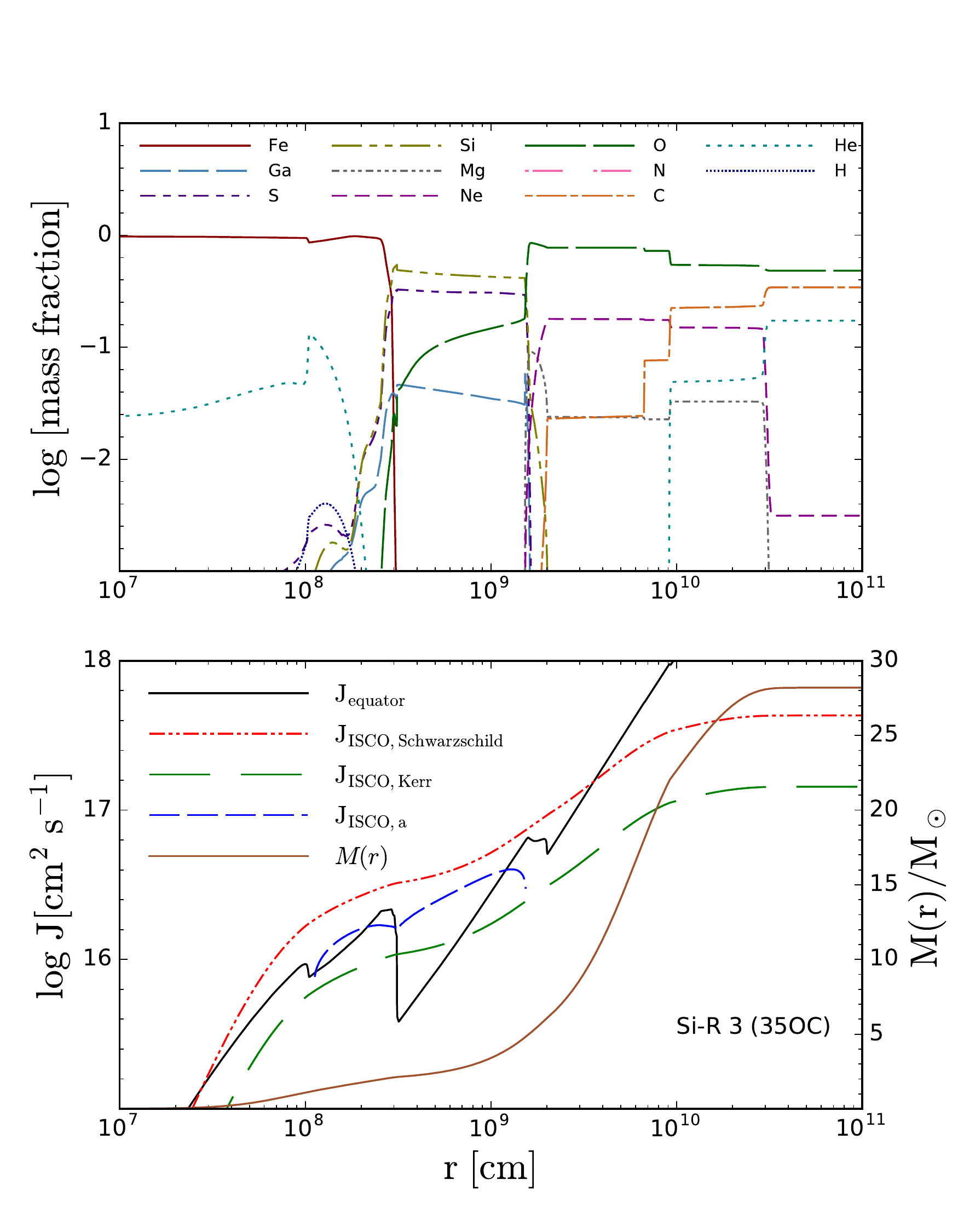}
\caption{Same as Fig.~\ref{fig:HE16F}, but for presupernova model Si-R 3 (35OC). The Si-R 3 (35OC) model are evolved from a rapidly rotating star with initial mass $35 M_\odot$ and $10\%$ solar metallicity. The final mass is $28.07 M_\odot$ considering a fraction of $0.1 \dot{M}$ of the standard WR mass-loss rate and the Fe core has an angular momentum at core collapse ${\mathcal J}_{\rm Fe~core} = 230 \times 10^{47}\rm~erg~s$ considering the effect of magnetic torques~\cite{Woosley:2005gy}. \label{fig:35OC}}
\end{figure}

\subsection{\label{sec:two_2}The composition of the accretion disk}
The composition of an accretion disk is determined by the falling stellar material. In this work, we adopt a simplified model where the influence from disk wind and accretion shock are ignored~\cite{Kumar:2008dv, Kumar:2008dr}. The free fall time scale for a fluid element at radius $r$ can be estimated to be $t_{\rm ff} \sim t_s(r) + (r^3/GM(r))^{1/2} \sim 2 (r^3/GM(r))^{1/2}$ where $t_s(r)$ is the time for the propagation of information of the core collapse. Considering the angular momentum conservation, only fluid elements whose specific angular momentum is $J(r) > J_{\rm ISCO}$ can form the accretion disk around the black hole, where $J_{\rm ISCO}$ is the specific angular momentum at the inner stable circular orbit (ISCO) $r_{\rm ISCO}$. The specific angular momentum in the progenitor stars at radius $r$ are estimated as $J(r) = r^2 {\rm sin}^2 {\rm \theta}  \ \Omega(r)$, where $\rm \theta$ is the angle to the polar axis. 
We show the distribution of nuclear mass fraction (upper panel) and specific angular momentum at the onset of core collapse for the five presupernova models (lower panel) in Fig.~\ref{fig:HE16F}-\ref{fig:35OC}. The kinks appeared on the curve of the distributions of specific angular momentum represent the transition region where their composition change. The specific angular momentum at ISCO, $J_{\rm ISCO}$, is determined by the black hole mass and spin parameter. 
In this work, we adopt the analytic formula of $J_{\rm ISCO}$ given by Ref.~\cite{Bardeen:1972fi},
\begin{equation}
\label{eq1}
J_{\rm ISCO} = \frac{2 G M_{\rm BH}}{3^{3/2} c} \left[1 + 2 \left(3 \frac{r_{\rm ISCO}}{r_g} - 2\right)^{1/2}\right],
\end{equation}
where $r_g = G M_{\rm BH} / c^2$ and 
\begin{equation}
r_{\rm ISCO} = \frac{G M_{\rm BH}}{c^2} \{ 3 + z_2 - [(3 - z_1)(3 + z_1 + 2 z_2]^{1/2}\},
\end{equation}
with
\begin{equation}
z_1 = 1 + (1 - a_{\rm BH}^2)^{1/3} [(1 + a_{\rm BH})^{1/3} + (1 - a_{\rm BH})^{1/3}],
\end{equation}
and
\begin{equation}
z_2 = (3 a_{\rm BH}^2 + z_1^2)^{1/2}.
\end{equation}
For simplicity, we assume that the black hole mass is equal to the enclosed mass within radius $r$. The spin parameter can be estimated using the formula 
$a_{\rm BH} = c {\mathcal J}_{\rm BH} / G M_{\rm BH}^2 \sim c  {\mathcal J}(r) / G M(r)^2$, where ${\mathcal J}(r)$ is the enclosed angular momentum and $M(r)$ is the enclosed mass. We show the distribution of $J_{\rm ISCO, a_{\rm BH}}$ as blue lines for the five presupernova models in the lower panel of Fig.~\ref{fig:HE16F}-\ref{fig:35OC}.
We also show $J_{\rm ISCO}$ for two extreme cases: a Schwarzschild black hole ($a_{\rm BH} = 0$, red lines) or an extreme Kerr black hole ($a_{\rm BH} = 1$, green lines). 

We can see that the inner part of progenitor stars does not have enough angular momenta to form an accretion disk.  The critical radius $r_c$ in the progenitors where the stellar matter begins to form the accretion disk can be estimated under the condition $J_{\rm ISCO, a_{\rm BH}} = {\mathcal J}(r)$. 
Taking presupernova model 16TI in Ref.~\cite{Woosley:2005gy} as an example, a $\sim 3.3 M_\odot$ material in the inner core will collapse into a black hole promptly. We find that $r_c$ has the typical value $\sim 2 \times 10^9 \ \rm cm$ with $\theta = \pi/2$ and the related fall back time initiating from the radius $r_c$ is $t_c \sim 4.3 \rm~s$. 
We assume that the jet nuclear component reflects the composition of the newly formed accretion disk at time $t_c$, and the matter that composes the accretion disk at time $t_c$ should come from stellar material at the radius $r_c$.
We tabulate the values of the critical radius $r_c$, enclosed mass $M_c(r_c)$ and the jet nuclear composition for the five presupernova models in Table~\ref{tab:table1}. We classify our nuclear composition models into two groups: Si-free and Si-rich. The Si-free models include Si-F 1 (HE16F) and Si-F 2 (16TI) which have very little fraction of nuclei heavier than oxygen nuclei, while the Si-rich models include Si-R 1(12TJ), Si-R 2 (16TJ) and Si-R 3 (35OC) which have a significant fraction of nuclei heavier than oxygen nuclei, especially the silicon nuclei.

Note that the nuclei in the inner disk region may be photodisintegrated into free nucleons if the temperature and density are high enough~\cite{MacFadyen:1998vz}. This is the case for the neutrino-dominated accretion flow (NDAF) where neutrino annihilation drives the relativistic outflow, and the disk should be composed of lighter composition dominated by protons once the quasi-statistical equilibrium (QSE) is established with temperature $T \gtrsim 4 \times 10^9 {\rm~K}$~\cite{Shibata:2015mva}. 
Thus, in the classical fireball model, it has been shown that the jet composition of HL GRBs is likely to be dominated by protons and neutrons (see Ref.~\cite{Horiuchi:2012by} and references therein).
However, we can expect the survival of nuclei at the outer regions or at the surface of the accretion disk for LL GRBs or magnetic energy dominated GRBs where the temperature and density can be lower.

\subsection{\label{sec:two_3}The fate of nuclei during initial loading}
Nuclei will be destroyed due to collisions with target protons or photons which have energy larger than the nuclear binding energy $\sim~10\rm~MeV$ in the nuclear rest frame. The temperature of the thermal photons at the jet base can be estimated using the formula $a_{\rm SB} T_0^4 = L_{\rm rad, 0}/\Omega_0 r_0^2 \Gamma_0^2 c$, where $T_0$ is the temperature, $L_{\rm rad, 0}$ is the jet initial radiation luminosity, $\Omega_0$ is the jet opening angle, $r_0$ is the location of jet base, $\Gamma_0$ is the jet initial Lorentz factor, and $a_{\rm SB}$ is the radiation constant. The temperature have value $T_0 \simeq 1.3 \ {\rm MeV} \ L_{\rm rad, 48}^{1/4} r_{0, 8}^{-1/2} \Omega_{0, -1}^{-1/4} \Gamma_0^{-1/2}$ which means that nuclei can be photodisintegrated into free nucleons by photons at the high-energy tail~\cite{Horiuchi:2012by}. 
The optical depth of nuclei at the jet base can be estimated using the giant-dipole resonance (GDR) approximation. In this work, we consider both the optical depth $\tau_{A\gamma} \approx n(\varepsilon) \sigma_{A\gamma} r_0 / \Gamma_0$ and effective optical depth $f_{A\gamma} \approx n(\varepsilon) \sigma_{A\gamma} \kappa_{A\gamma} r_0 / \Gamma_0$ where $n(\varepsilon)$ is the photon number density related  to the GDR process, $\sigma_{A\gamma}$ is the GDR cross section,  and $\kappa_{A\gamma}$ is the inelasticity~\cite{Zhang:2017hom}. The optical depth, $\tau_{A\gamma}$, represents the interaction rate of nuclei with target photons, while the effective optical depth, $f_{A\gamma}$, indicates the energy loss efficiency of the nuclei. These two values are different, since the energy change rate in each interaction can be small.

\begin{figure}
\includegraphics[width=\linewidth]{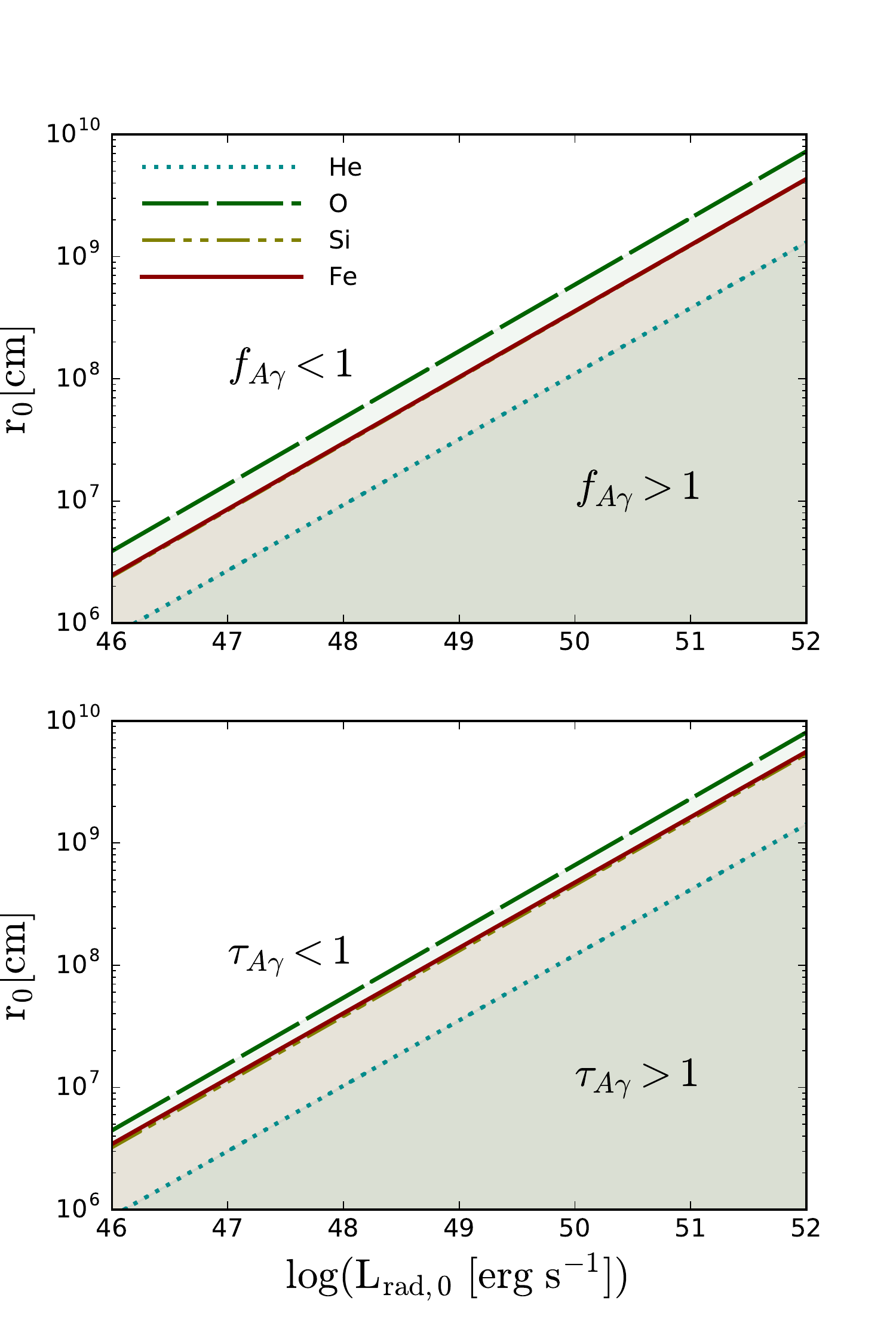}
\caption{Constraints on the initial radiation luminosity $L_{\rm rad, 0}$  and the location of jet base $r_0$ for the survival of nuclei with various chemical species: He, O, Si, and Fe during the initial loading at jet base. The blank (shaded) region beyond (below) the iron curve has an optical depth $\tau_{A\gamma} < 1$ ($\tau_{A\gamma} >1$) in the lower panel and effective optical depth $f_{A\gamma} > 1$ ($f_{A\gamma} > 1$) in the upper panel. The optical depth $\tau_{A\gamma}$
represents the interaction rate, while the effective optical depth $f_{A\gamma}$ indicates the energy loss efficiency of nuclei. \label{fig:fate_loading_plot}}
\end{figure}

We estimate the values of $f_{A\gamma}$ (upper panel) and $\tau_{A\gamma}$ (lower panel) as functions of the location of jet base $r_0$ and initial radiation luminosity $L_{\rm rad, 0}$, as have been shown in Fig.~\ref{fig:fate_loading_plot}. We can see that nuclei are more easier to survive in the upper left region where the initial radiation luminosity $L_{\rm rad, 0}$ is lower and the jet base is further from the black hole. A reasonable assumption of the location of jet base is that it has a similar scale as the critical radius of disk formation, $r_{\rm c} \sim 10^9 \rm~cm$. Thus, it is reasonable to ignore the photodisintegration of nuclei during the initial loading at the jet base for LL GRBs.

\subsection{\label{sec:two_4} Hypernova ejecta composition}
Motivated by the connection between hypernova and LL GRBs, we also consider the hypernova ejecta composition where the heavy nuclei component are synthesized during the expansion of the semi-relativistic ejecta. This is possible if the ejecta is launched a bit earlier than the jet and the synthesized material is entrained onto the jet later. The ejecta may be the wind from accretion disk~\cite{MacFadyen:1998vz} or the shocked stellar material driven by the relativistic outflow~\cite{Barnes:2017hrw}. 

The optical counterpart of GRB 980425 is an energetic Type \RN{1}c supernova SN 1998bw which has a larger kinetic energy ($\sim 10^{52}\rm~erg$) compared to usual supernovae. In this work, we adopt the yields of hypernova ejecta model CO138E50 calculated by Ref.~\cite{Nakamura:2000ms}, which can reproduce the early light curve of SN 1998bw. We show the synthesized nuclear mass fraction in Table~\ref{tab:table1}. Compared to the initial loading scenario, the hypernova ejecta model contains a large fraction of iron-group nuclei with $X_{\rm Fe} \simeq 0.09$.

\section{\label{sec:three}The origin of UHECR nuclei in LL GRBs}
\subsection{Acceleration and survival}
It has been suggested that nuclei in the relativistic outflow can be accelerated to ultrahigh energies in the dissipation region of GRBs where the internal and/or external shock taking place~\cite{Waxman:1995vg, Vietri:1995hs}. The relevant acceleration mechanism is diffusive shock acceleration (or Fermi acceleration)~\cite{Bell:1978zc, Blandford:1987pw, Sironi:2015oza}, and the first-order Fermi acceleration mechanism is wildly used where nuclei acquire energy when bouncing back and forth around the shock front. We rely on the results of the test-particle approximation, where the shock structure is unaffected by the accelerated particles. The acceleration time scale can be estimated as $t_{\rm acc} = \eta r_L /c$, where $r_L = E_A / Z e B $ is the Larmor radius for particles of energy $E_A$ and charge $Z$, and $B$ is the magnetic field strength measured in the jet comoving frame. The parameter $\eta$ represents the acceleration efficiency whose typical value is $\eta > 1$~\cite{Murase:2006mm}.

The internal shock model is one of the most popular models to explain the prompt emissions of GRBs. In this model, the faster moving ejecta catches the slower ones and produce a shock wave~\cite{Rees:1994nw, Meszaros:2006rc}. The radius of shock production can be estimated using time variability $R \simeq 2 \Gamma^2 \delta t \ c \simeq 1.2 \times 10^{15}~\Gamma_{1}^2 (\delta t / 200~{\rm s})\rm~cm$, where $\delta t$ is the variability time scale~\cite{Murase:2008mr}. The comoving magnetic field strength in the internal shock region can be estimated to be $B \simeq220~\xi_{B}^{1/2} L_{\gamma\rm iso, 47}^{1/2} R_{15}^{-1}  \Gamma_{1}^{-1} \rm~G$ assuming that a fraction of the outflow kinetic energy is converted into the magnetic energy. The observed maximum acceleration energy can be estimated under the condition $t_{\rm acc} \leq t_{\rm dyn}$ where $t_{\rm acc}$ is the acceleration time scale and $t_{\rm dyn} \equiv R / \Gamma \beta c$ is the dynamical time scale, and we have $E_{A, \rm dyn} = \Gamma \eta^{-1} Z e B (R/\Gamma) \simeq 10^{18.7} (\eta/15)^{-1}Z \xi_{B}^{1/2} L_{\gamma\rm iso, 47}^{1/2} \Gamma_{1}^{-2} \rm~eV$ measured in the observer frame. The maximum CR energy that limited by synchrotron energy losses is $E_{A, \rm syn} \simeq3.2\times10^{19} \ A^2 Z^{-3/2} (\eta/15)^{-1/2} \xi_{B}^{-1/4} L_{\gamma\rm iso,47}^{-1/4} \Gamma_{1}^{3/2} R_{15}^{1/2} \rm~eV$ in the observer frame. 
If we introduce the escape boundary that may be caused by the region where a strong magnetic field amplification is expected, for the comoving size $l_{\rm esc}\approx x_{\rm esc}(R/\Gamma)$, we have $E_{A, \rm esc}\simeq 10^{18.2} (\eta/15)^{-1}Z \xi_{B}^{1/2} x_{\rm esc,-0.5}L_{\gamma\rm iso, 47}^{1/2} \Gamma_{1}^{-2} \rm~eV$~\cite{Ohira2010AA}.

\begin{figure}
\includegraphics[width=\linewidth]{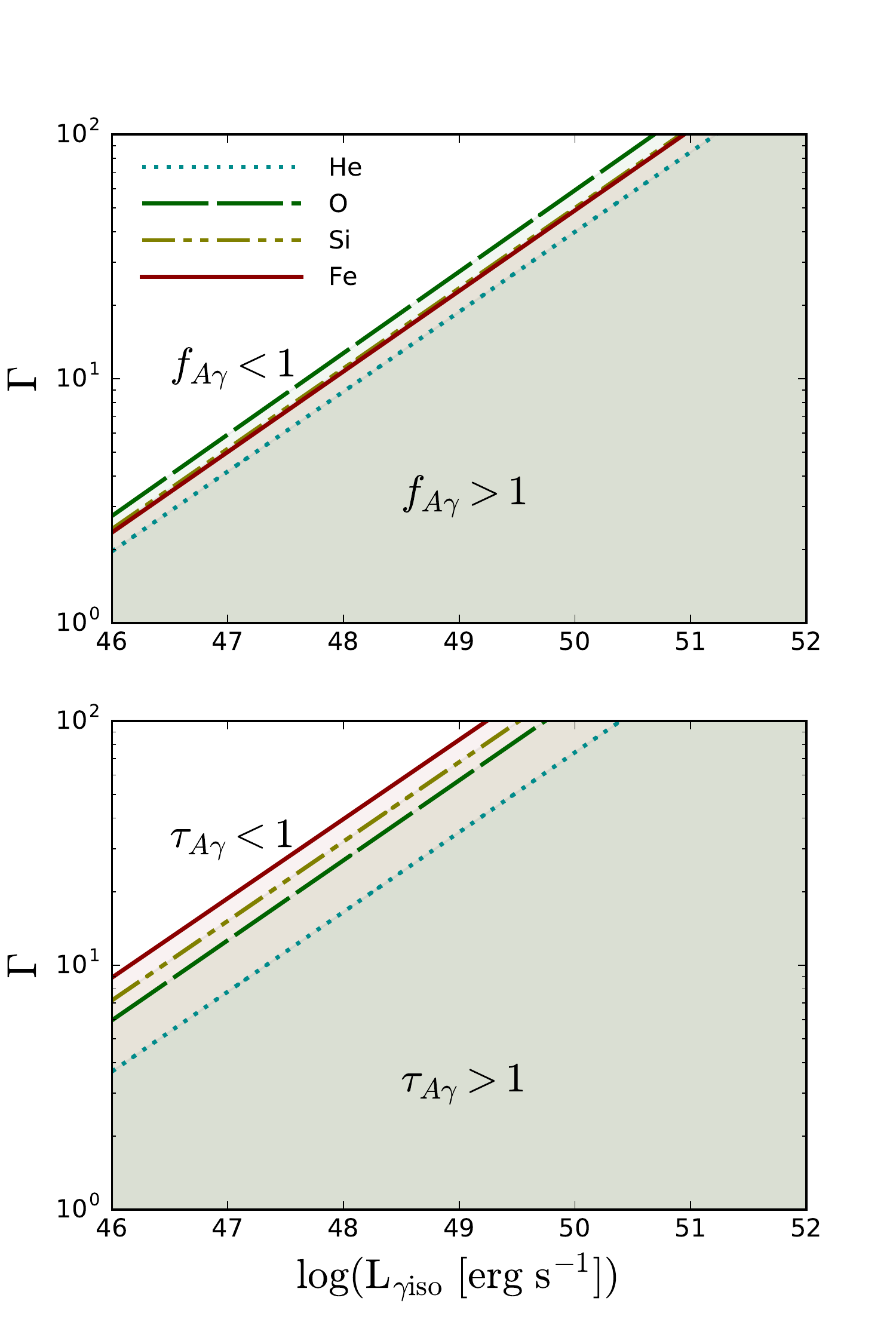}
\caption{Constraints on the isotropic equivalent radiation luminosity $L_{\gamma\rm iso}$ and Lorentz factor $\Gamma$ for nuclei with various chemical species with energy $E_A = 10^{20}\rm~eV$ to survive in the internal shock region. The upper panel corresponds to the energy loss time scales and the lower panel corresponds to the interaction time scales. We assume the radius where internal shock taking place is $R \simeq 1.2 \times 10^{15} \rm~cm$ and the break energy of prompt emission is $\varepsilon_b = 500\rm~eV$ in the jet comving frame. \label{fig:fate_IS_plot}}
\end{figure}

Now, following Ref.~\cite{Murase:2008mr}, let us assess the survivability of the UHECR nuclei in the internal shock region. We model the GRBs prompt emission as a broken power-law
\begin{equation}
\frac{dn}{d\varepsilon} = \frac{(L_{\gamma\rm iso} / 5) e^{-\varepsilon/\varepsilon_{\rm max}}}{4\pi r^2 \Gamma^2 c {\varepsilon_b}^2} \left\{ \begin{array}{ll} (\varepsilon / \varepsilon_b)^{-1} & (\varepsilon_{\rm min} \leq \varepsilon < \varepsilon_b) \\ (\varepsilon / \varepsilon_b)^{-2.2} & (\varepsilon_b \leq \varepsilon \leq \varepsilon_{\rm max}) \end{array}, \right.
\end{equation}
where $\approx L_{\gamma\rm iso}/5$ is the photon luminosity at the break energy $\varepsilon_b$ and  $\varepsilon_{\rm min} = 1 \rm~eV$, $\varepsilon_{\rm max} = 10 \rm~MeV$ \cite{Murase:2008mr, Horiuchi:2012by}. Similar to the discussion in Sec.~\ref{sec:two_3}, we consider two kinds of time scales: the interaction time scale which is related to the optical depth $\tau_{A\gamma} \equiv t_{A\gamma-\rm int}^{-1} / t_{\rm dyn}^{-1}$ and the energy loss time scale which corresponds to the effective optical depth $f_{A\gamma} \equiv t_{A\gamma}^{-1} / t_{\rm dyn}^{-1}$; see the details in Refs.~\cite{Murase:2008mr, Zhang:2017hom}.
We estimate the value of $f_{A\gamma}$ (upper panel) and $\tau_{A\gamma}$ (lower panel) as functions of the kinetic energy and Lorentz factor of the relativistic outflow. In Fig.~\ref{fig:fate_IS_plot}, we can see that nuclei are easier to survive for dimmer GRBs.

\begin{figure}
\includegraphics[width=\linewidth]{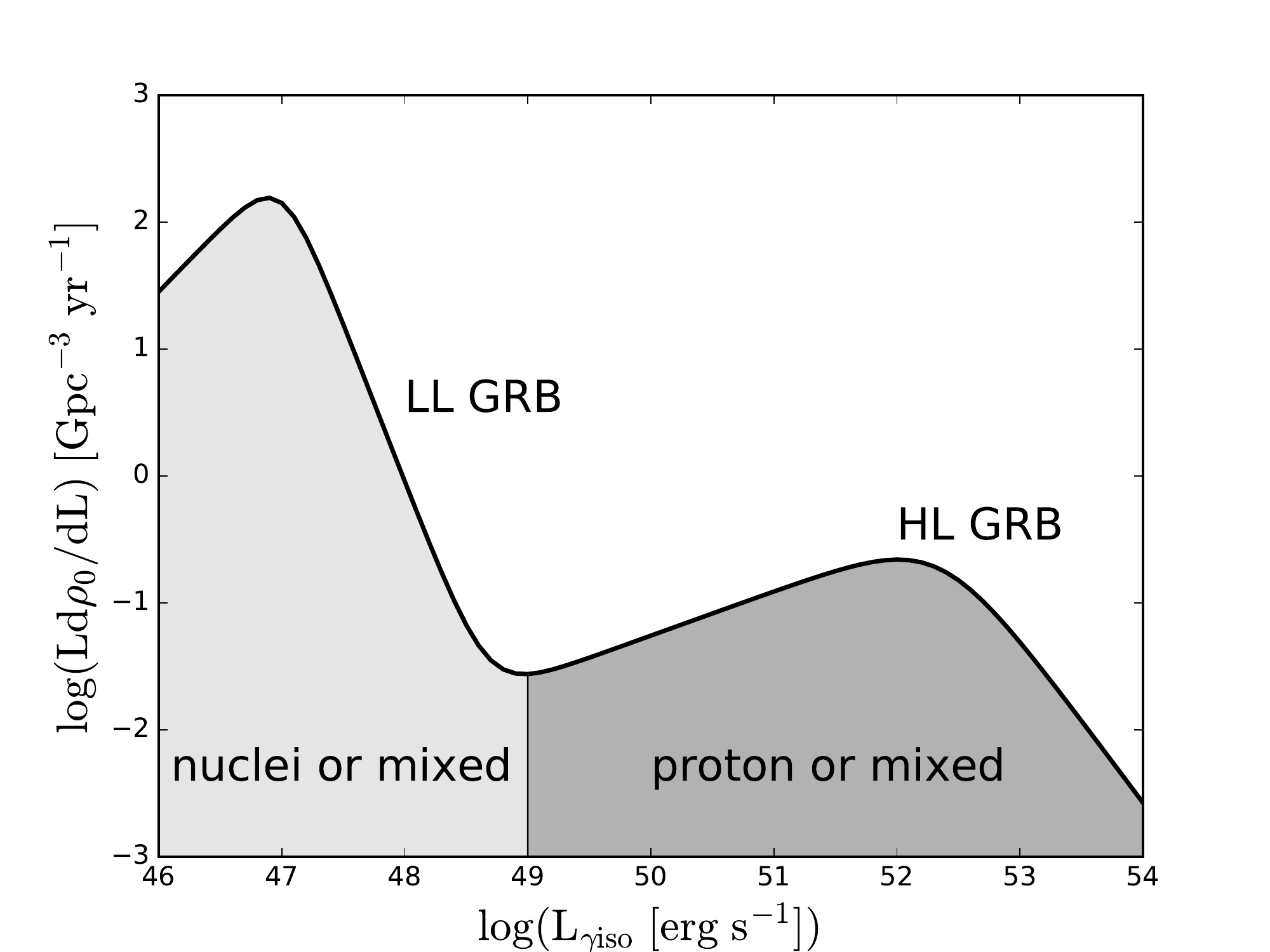}
\caption{The GRB luminosity function used in this work. The parameters for LL GRBs are $L_{\rm min}^{\rm LL} = 10^{46} \rm~erg~s^{-1}$, $L_{\rm max}^{\rm LL} = 10^{49} \rm~erg~s^{-1}$, $L_b^{\rm LL} = 10^{47} \rm~erg~s^{-1}$, $\alpha_1^{\rm LL} = 0.0$, $\alpha_2^{\rm HL} = 3.5$, and the parameters for HL GRBs are $L_{\rm min}^{\rm HL} = 10^{49} \rm~erg~s^{-1}$, $L_{\rm max}^{\rm HL} = 10^{54} \rm~erg~s^{-1}$, $L_b^{\rm HL} = 10^{52.35} \rm~erg~s^{-1}$, $\alpha_1^{\rm HL} = 0.65$, $\alpha_2^{\rm HL} = 2.3$ \cite{Liang:2006ci}. Schematically we indicate that the UHECRs in LL GRBs have a composition dominated by nuclei, while HL GRBs are likely to have a proton-rich or mixed composition. \label{fig:fate_LF}}
\end{figure}

The luminosity function of both LL GRBs and HL GRBs can be described using the following formula~\cite{Liang:2006ci},
\begin{equation}
\frac{d\rho_0}{dL} = A_0 \left[\left(\frac{L}{L_b}\right)^{\alpha_1} + \left(\frac{L}{L_b}\right)^{\alpha_2}\right]^{-1},
\end{equation}
where $A_0$ is the normalization parameter, as have been shown in Fig.~\ref{fig:fate_LF}. The LL GRBs are defined to have isotropic radiation luminosity $L_{\gamma\rm iso} \leq 10^{49}\rm~erg~s^{-1}$, where nuclei can survive. With this luminosity function, the contribution from LL GRBs is more important in the local universe. Note that the durations of GRB 060218 and GRB 100316D were $\sim{10}^3-{10}^4$~s, which is about 100 times longer than the typical duration of HL GRBs, $\sim10-100$~s. Although the LL GRB rate is highly uncertain, such a setup may be justified if the LL GRB outflows are more contaminated by baryons (i.e. if they originate from a so-called dirty fireball). In this work, we mainly focus on the connection between LL GRBs and UHECR nuclei. 

Although the LL GRB contribution is more important in our setup, it is possible to discuss the HL GRB contribution. The HL GRBs have higher luminosities, shorter variability times, and higher Lorentz factors. In the context of the Auger data, the survivability of heavy nuclei was first studied by Refs.~\citep{Murase:2008mr, Wang:2007xj}. For the fiducial Lorentz factor, $\Gamma\sim300$, the nuclei are completely destroyed for internal shock radii of $\lesssim{10}^{14}$~cm. For the fiducial luminosity of $L_{\gamma\rm iso}\sim10^{52}~{\rm erg}~{\rm s}^{-1}$, the nucleus-survival is shown to be possible only at sufficiently large radii of $\sim{10}^{15}$~cm, and the photodisintegration occurs in the partial survival regime, i.e. $\tau_{A\gamma}\gtrsim1\gtrsim f_{A\gamma}$. Such special cases are studied in Refs.~\cite{Globus:2014fka, Biehl:2017zlw}, but parameters need to be tuned to fit the UHECR spectrum and composition. This is especially case if the luminosity function and/or multizone effect are considered. 
At large emission radii allowing the nucleus survival, the internal shock model suffers from an issue in explaining the peak energy by synchrotron emission. Magnetic fields may be too weak to be consistent with the synchrotron peak energy, so that the number of accelerated electrons needs to be small enough to boost the electron injection Lorentz factor (e.g.,~\cite{Bosnjak:2008bd}). 
Later, we discuss the effect of the HL contribution assuming the proton composition. This corresponds to the assumption that the nuclei are completely destroyed somewhere. 

\subsection{Escape and injection spectrum}
The accelerated nuclei usually follow a power-law distribution with a spectral index, $s_{\rm acc}\sim2.0-2.2$, as a prediction of the diffusive shock acceleration mechanism (e.g.,~\cite{Keshet:2004ch,Sironi:2010rb,Caprioli:2014tva}). However, in general, the spectrum of UHECR nuclei ejected from GRBs does not need to be equal to the acceleration spectrum (see, e.g.,~\cite{Zhang:2017hom} for the case of tidal disruption events). 
In fact, a harder injection spectrum is achieved if we take into account escape processes, such as the simple direct escape model from the emission region (that is mainly the downstream)~\cite{Baerwald:2013pu, Zhang:2017hom, Heinze:2015hhp, Biehl:2017zlw} and the neutron escape model (e.g., \cite{Mannheim:1998wp,Dermer:2012rg}). 
Here we consider the diffusive shock acceleration, in which particle escape from the escape boundary in the upstream has been theoretically calculated and it has been applied to supernova remnants and radio galaxies~\cite{Ohira2010AA, Atoyan:2002gu, Caprioli:2009bf, Drury:2010am, Katz:2010tv}. 
In the escape-limited model, only cosmic rays whose energy is close to the maximum acceleration energy can escape from the sources, and the spectrum of escaping particles at given time is essentially approximated by a delta function. In this work, for simplicity, we extrapolate the results of the non-relativistic diffusive shock acceleration theory, where we use the following expression for the spectrum of escaping cosmic rays with $s_{\rm acc}=2$ as the injection spectrum of UHECRs,
\begin{equation}
\label{eq:escape}
\frac{dN_{A^\prime}}{dE^\prime} = f_{A^\prime} N_0 {\rm exp} \left[ - {\rm ln}^2 \left(\frac{E^\prime}{Z E_{p, {\rm max}}^\prime}\right) \right],
\end{equation}
where $f_{A^\prime}$ is the number fraction of nuclei with mass number $A^\prime$, $N_0$ is the normalization parameter which depends on the radiation luminosity and $Z E_{p, {\rm max}}^{\prime}$ is the CR maximum acceleration energy. Note that the instantaneous spectrum of CRs escaping from the sources does not have to be a power-law function that is used in most of the previous works. However, for the completeness, we also considered cases with power-law spectra, and we find that the Auger data can also be well fitted by a hard power-law spectral index. See the next section for details. Note that in this work we do not specify the UHECR acceleration site, which can be internal shocks or an external (reverse) shock. The external shock model naturally provides an interesting possibility to explain the PeV neutrino flux simultaneously.

\section{\label{sec:four}Propagation and results}
\subsection{Propagation}
We numerically propagate UHECR nuclei using the publicly available Monte Carlo code CRPropa 3~\cite{Batista:2016yrx}. CRPropa is one of the state-of-the-art numerical simulation frameworks that enable us to propagate UHECR nuclei in the intergalactic space taking into account various energy loss processes, such as the photomeson production process for protons and neutrons, the photodisintegration process for heavy nuclei, the Bethe-Heitler process which can be treated as continuous energy loss process, and adiabatic energy losses due to the expansion of the universe~\cite{Kampert:2012fi}. The background photons are mainly composed of the CMB and extragalactic background light (EBL) photons, and the latter is important for the photodisintegration process of UHECR nuclei in the lower energy range. In this work, we adopt a semi-analytic EBL model by Ref.~\cite{Gilmore2012}. 

The observed flux of UHECR nuclei with $A$ at Earth can be calculated using the following formula,
\begin{eqnarray}
\Phi_A(E) &=& \sum_{A^\prime} \frac{c}{4\pi} \int_{z_{\rm min}}^{z_{\rm max}} dz  \left| \frac{dt}{dz}\right| F_{\rm GRB}(z) \nonumber \\ &\times & \int_{L_{\rm min}}^{L_{\rm max}} \frac{d \rho_0}{dL} \int_{E^\prime_{\rm min}}^{E^\prime_{\rm max}} dE^\prime \frac{dN_{A^\prime}}{dE^\prime}\frac{d\eta_{A A^\prime}(E, E^\prime,z)}{dE},\,\,\,\,\,\,\,\,\,\,
\end{eqnarray}
where $F_{\rm GRB}(z)$ is the redshift distribution parameter of long GRBs which trace the star formation history (SFH)~\cite{Sun:2015bda}, $\rho_0$ is the local event rate of GRBs, $d\rho_0/dL$ is the GRB luminosity function in the local universe~\cite{Liang:2006ci}, and $\eta_{A A^\prime}(E, E^\prime,z)$ is the fraction of generated cosmic rays of mass $A$ and energy $E$ from parent particles of mass $A^\prime$ and energy $E^\prime$~\cite{Zhang:2017hom}. The minimum redshift of the sources is set to $z_{\rm min} = 0.00055$ (i.e. $d=2.4$~Mpc), which is slightly smaller than that of the nearest starburst galaxy (SBG), NGC 253. But the results are not affected even if the minimum distance is set to that to another SBG, M82. As described in Section~\ref{sec:three}, the maximum energy and nucleus survival are evaluated following the procedure used in Refs.~\cite{Murase:2008mr,Horiuchi:2012by}.
Then, as well as the photodisintegration cross section data, we use the same method as in Ref.~\cite{Zhang:2017hom} to calculate the final spectrum and the distribution of $\left\langle X_{\rm max}\right\rangle$ and $\sigma(X_{\rm max})$~\cite{Aab:2017tyv}. For the fitting purpose, we fix the nuclear maximum acceleration energy to be $Z E_{p, {\rm max}}^\prime = 10^{18.2} Z {L_{\gamma\rm iso, 47}}^{1/2} \rm~eV$ as a simplified method. We show the best-fit result of the $\chi^2$ analysis, considering the effect of the systematic uncertainty on the measured CR energy scale $\sigma_E = 14\%$~\cite{Aab:2017njo}, and using $\delta_E$ as a free parameter to account for such uncertainty which is defined as $E = E^{\rm Auger} (1 + \delta_E)$~\cite{Heinze:2015hhp}. 

\begin{figure}
\includegraphics[width=\linewidth]{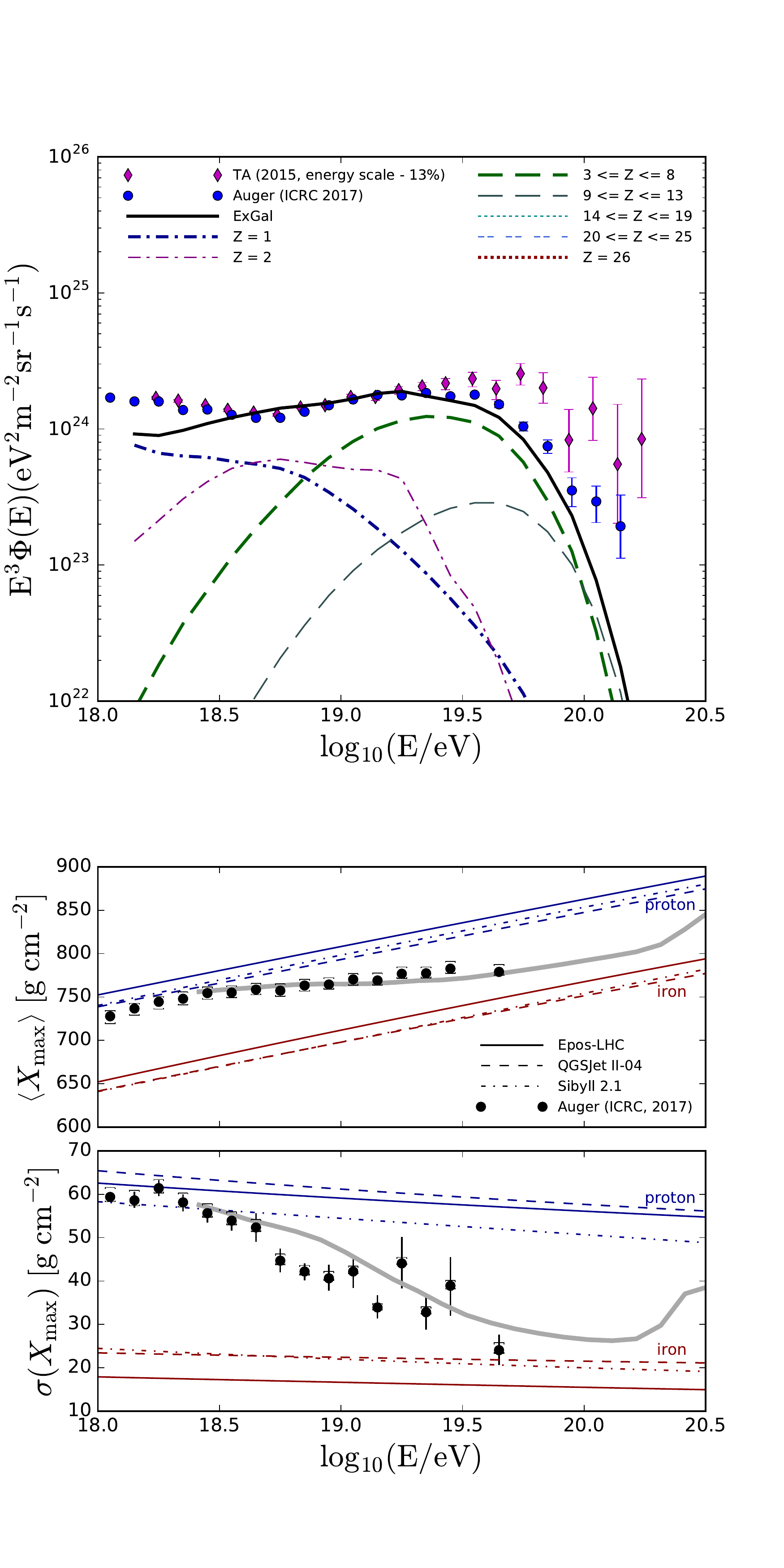}
\caption{The UHECR nuclei spectrum and distribution of $\left\langle X_{\rm max}\right\rangle $ and $\sigma(X_{\rm max})$ calculated from model Si-F 1. The blue data points are taken from Auger~\cite{Aab:2017njo} and we also show the magenta data points measured by TA for comparison~\cite{Fukushima:2015bza}. The maximum acceleration energy is 
$Z E_{p, {\rm max}}^\prime = 10^{18.2} Z {L_{\gamma\rm iso, 47}}^{1/2} \rm~eV$ and $\delta_E = -0.14 $. 
\label{fig:HE16F_luminosity_results}}
\end{figure}

\begin{figure}
\includegraphics[width=\linewidth]{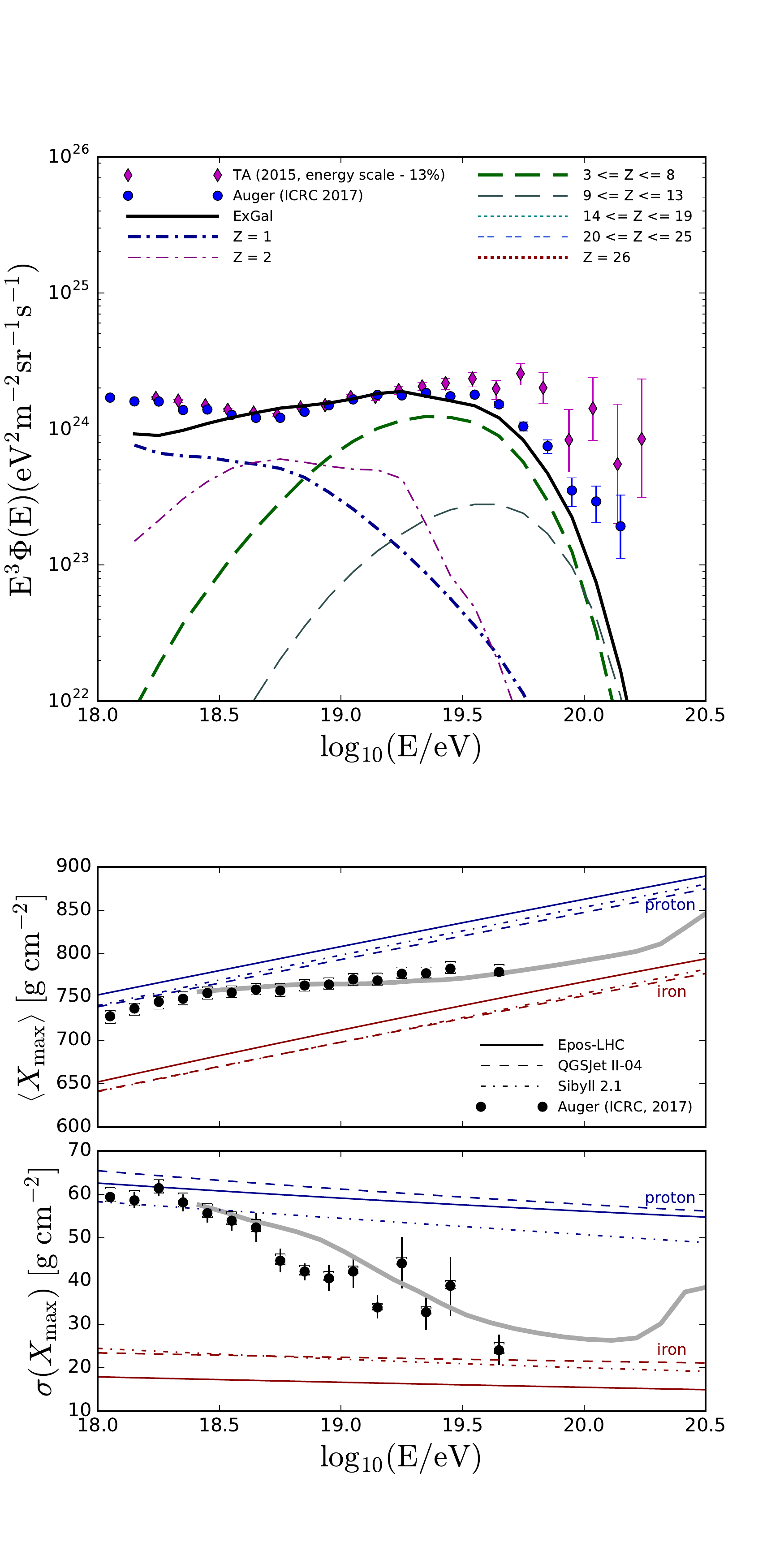}
\caption{Same as Fig.~\ref{fig:HE16F_luminosity_results} but for model Si-F 2. The maximum acceleration energy is $Z E_{p, {\rm max}}^\prime = 10^{18.2} Z {L_{\gamma\rm iso, 47}}^{1/2} \rm~eV$ and $\delta_E = -0.14$. 
\label{fig:16TI_luminosity_results}}
\end{figure}

\begin{figure}
\includegraphics[width=\linewidth]{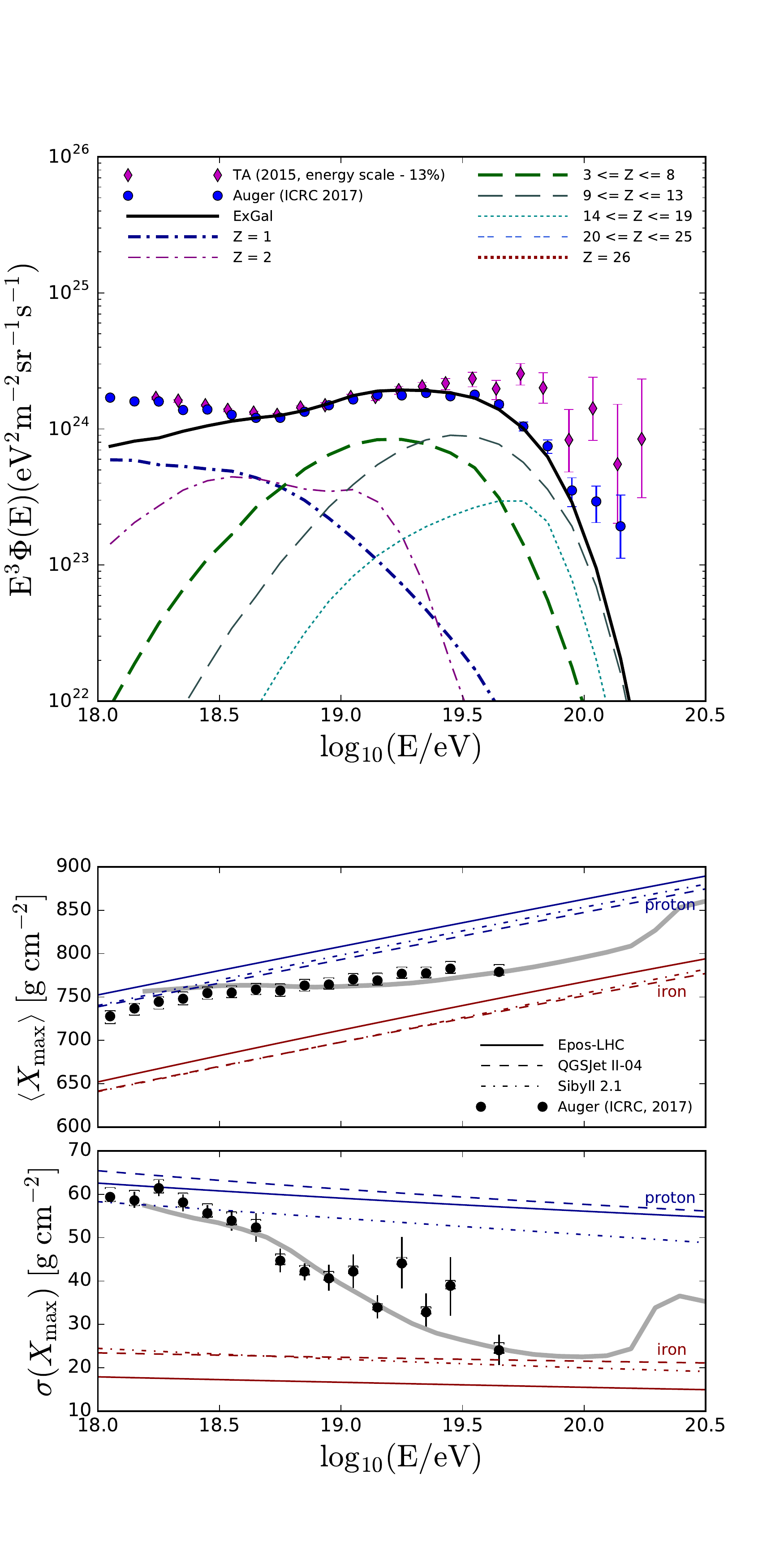}
\caption{Same as Fig.~\ref{fig:HE16F_luminosity_results} but for model Si-R 1. The maximum acceleration energy is $Z E_{p, {\rm max}}^\prime = 10^{18.2} Z {L_{\gamma\rm iso, 47}}^{1/2} \rm~eV$ and $\delta_E = 0.14$.
\label{fig:12TJ_luminosity_results}}
\end{figure}

\begin{figure}
\includegraphics[width=\linewidth]{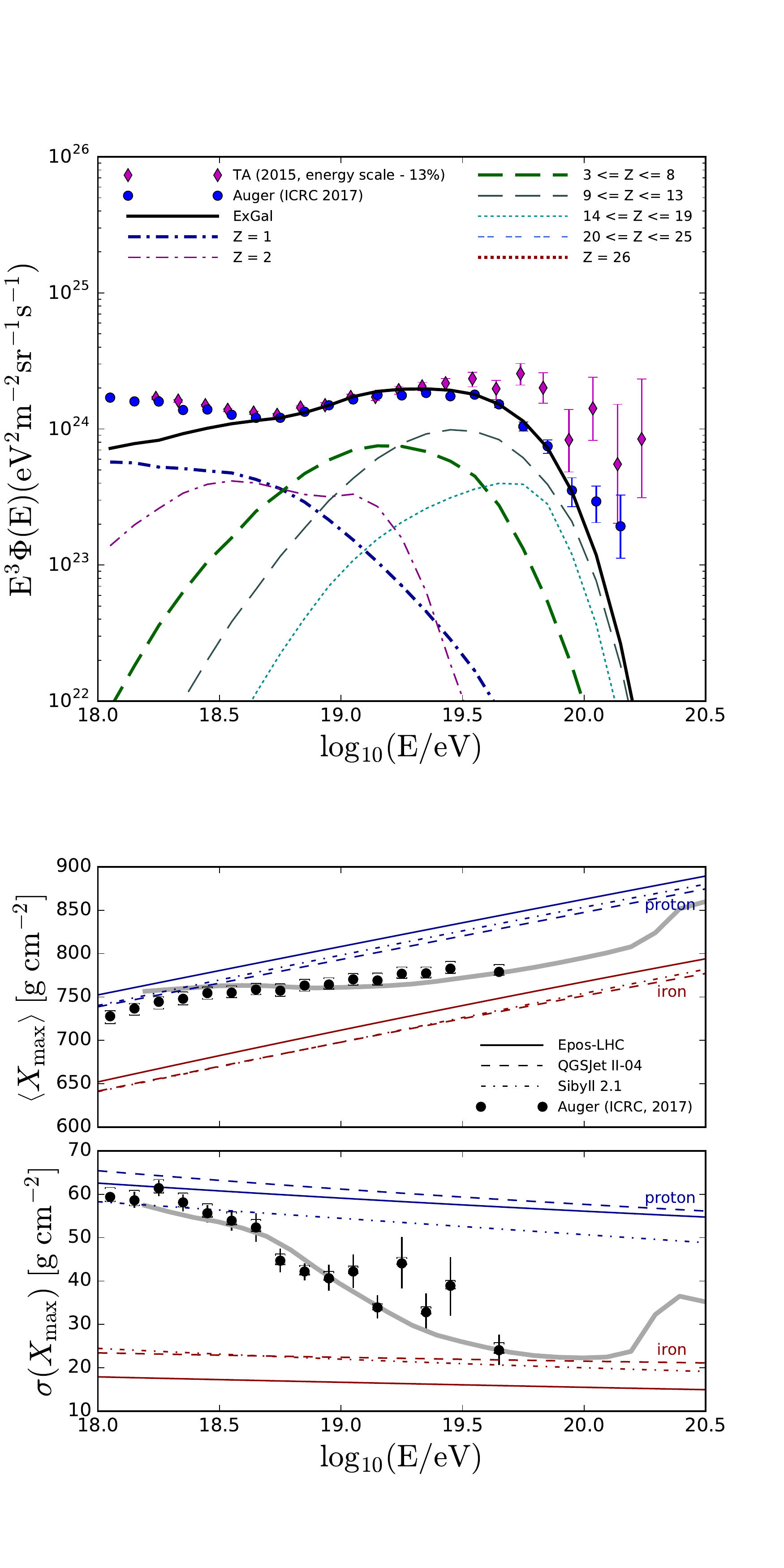}
\caption{Same as Fig.~\ref{fig:HE16F_luminosity_results} but for model Si-R 2. The maximum acceleration energy is $Z E_{p, {\rm max}}^\prime = 10^{18.2} Z {L_{\gamma\rm iso, 47}}^{1/2} \rm~eV$ and $\delta_E = 0.14$. 
\label{fig:16TJ_luminosity_results}}
\end{figure}

\begin{figure}
\includegraphics[width=\linewidth]{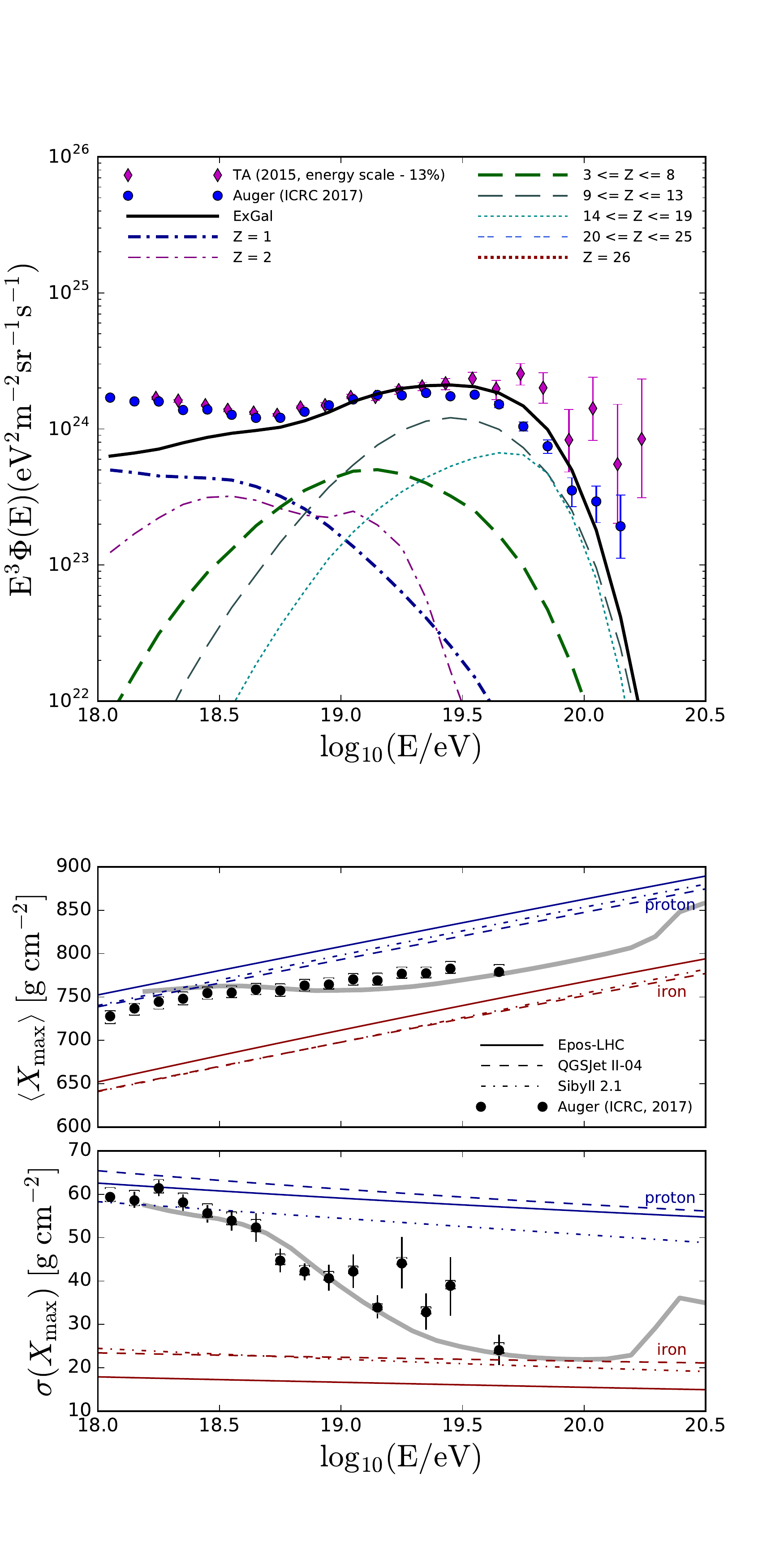}
\caption{Same as Fig.~\ref{fig:HE16F_luminosity_results} but for model Si-R 3. The maximum acceleration energy is $Z E_{p, {\rm max}}^\prime = 10^{18.2} Z {L_{\gamma\rm iso, 47}}^{1/2} \rm~eV$ and $\delta_E = 0.14$. 
\label{fig:35OC_luminosity_results}}
\end{figure}

\begin{figure}
\includegraphics[width=\linewidth]{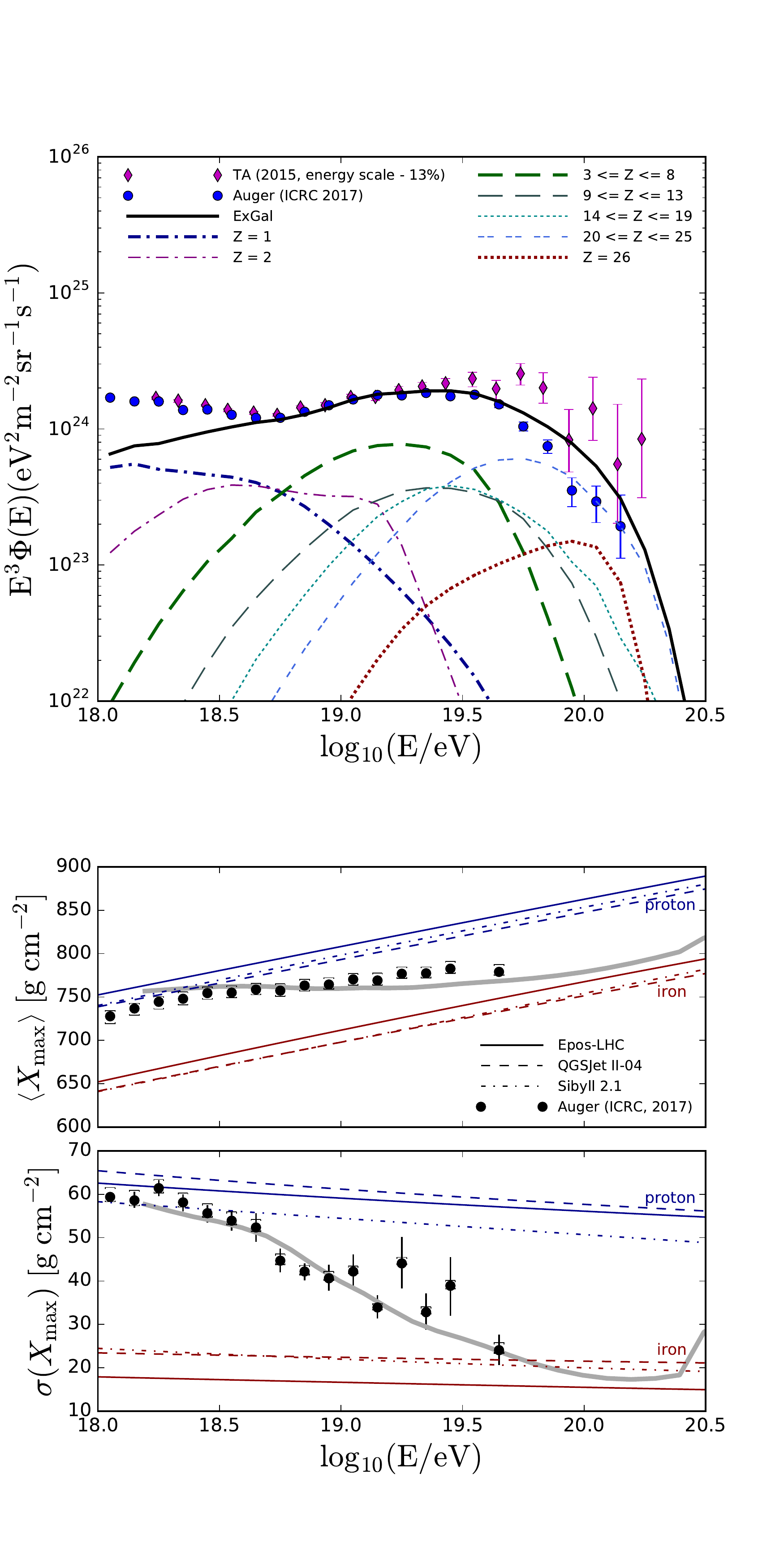}
\caption{Same as Fig.~\ref{fig:HE16F_luminosity_results} but for hypernova model. The maximum acceleration energy is $Z E_{p, {\rm max}}^\prime = 10^{18.2} Z {L_{\gamma\rm iso, 47}}^{1/2} \rm~eV$ and $\delta_E = 0.14$. 
\label{fig:hypernova_luminosity_results}}
\end{figure} 

\subsection{Results}
\begin{table}[b]
\caption{\label{tab:table2}%
Summary of the fitting parameters.}
\begin{ruledtabular}
\begin{tabular}{lcdc}
\textrm{Models}&
\textrm{${{\rm~log_{10}}E}$\footnote{We use the injection spectrum of UHECRs in Eq.~\ref{eq:escape} and show the maximum acceleration energy of protons for $L_{\gamma\rm iso} = 10^{47}\rm~erg~s^{-1}$.}}&
\multicolumn{1}{c}{\textrm{$\delta_E$}\footnote{The free parameter $\delta_E$ that can vary within the 14\% uncertainty in the CR energy scale.}}&
\textrm{$Q_{\rm UHECR,44}$\footnote{The energy injection rate of UHECRs, $Q_{\rm UHECR,44}\equiv Q_{\rm UHECR}/(10^{44}\rm~erg~Mpc^{-3}~yr^{-1}).$}}\\
\colrule
Si-F 1 & 18.2 & -0.14 & 2.0\\
Si-F 2 & 18.2 & -0.14 & 2.0\\
Si-R 1 & 18.2 & 0.14 &  3.6\\
Si-R 2 & 18.2 & 0.14 &  4.0\\
Si-R 3 & 18.2 & 0.14 &  3.6\\
Hypernova & 18.2 & 0.14 & 3.9\\
\end{tabular}
\end{ruledtabular}
\end{table}

We show the final spectrum and the distribution of $\left\langle X_{\rm max}\right\rangle $ and $\sigma(X_{\rm max})$ for each model in Fig.~\ref{fig:HE16F_luminosity_results}-\ref{fig:hypernova_luminosity_results}. 

For the Si-free models (Si-F 1 and Si-F 2), we find that it is difficult to fit the UHECRs spectrum measured by Auger. See Figs.~\ref{fig:HE16F_luminosity_results}-\ref{fig:16TI_luminosity_results}.  On the other hand, the Si-free models match $\left\langle X_{\rm max}\right\rangle $ very well and roughly fit $\sigma(X_{\rm max})$ due to the large error bars. The dominant composition of the Si-free models is oxygen nuclei, corresponding to progenitors with lower angular momenta in their inner core. It is not helpful to improve the fitting via increasing the maximum acceleration energy of oxygen nuclei, since higher maximum energies lead to too many light composition secondaries due to the shorter attenuation length of oxygen nuclei with $\sim 30~\rm Mpc$.

The Si-rich models (Si-R 1, Si-R 2, Si-R 3) can account for both the spectrum and the distribution of $\left\langle X_{\rm max}\right\rangle $ and $\sigma(X_{\rm max})$ measured by Auger. See Figs.~\ref{fig:12TJ_luminosity_results}-\ref{fig:35OC_luminosity_results}. Compared to the Si-free models, the Si-rich models have a larger fraction of silicon nuclei $(\sim 0.3)$, as shown in Table~\ref{tab:table1}. The Si-rich models can be achieved when the progenitors have larger angular momenta in their inner cores, where a significant fraction of the silicon layer material forms the accretion disk. 

Also, we find that the hypernova ejecta composition model can roughly fit the Auger data. See Fig.~\ref{fig:hypernova_luminosity_results}. Compared to the Si-free models and Si-rich models, the hypernova ejecta contains a large fraction of nickel or iron nuclei. The iron component becomes important at the highest-energy part of the UHECRs spectrum, where the composition is unknown.

In the lower panel of Fig.~\ref{fig:HE16F_luminosity_results}-\ref{fig:hypernova_luminosity_results}, we can see that $\left\langle X_{\rm max}\right\rangle $ and $\sigma(X_{\rm max})$ changes to a lighter composition at the highest energies. 
We find that protons and light nuclei at these energies are mainly produced as secondaries due to the photodisintegration of heavy nuclei that originate from the nearest sources which are located within several megaparsecs around Earth. Thus, this feature will disappear for rare sources, which imply that the minimum distance is large. (See Refs.~\cite{Murase:2008mr,Murase:2008sa,Takami:2011nn} for discussions about the source density of transients including LL GRBs.)
The UHECR nuclei at the highest energies mainly come from LL GRBs with luminosity $\sim10^{49}\rm~erg~s^{-1}$, and the energy dependence of the photodisintegration cross section used in CRPropa 3 also appears.

The required UHECR luminosity density, $Q_{\rm UHECR}$, for the five models are in the range of $(2-4)\times 10^{44}\rm~erg~Mpc^{-3}~yr^{-1}$, as shown in Table \ref{tab:table2}. 
It can be approximated by $Q_{\rm UHECR} = \xi_{\rm CResc} {\mathcal E}_{\gamma\rm iso} \rho_0^{\rm LL}$ , where $\xi_{\rm CResc}$ is the CR loading factor defined for escaping CRs. 
We find $\xi_{\rm CResc}$ has a value in the range of $\sim(10-20)~{\mathcal E}_{\gamma\rm iso, 50}$, assuming the local event rate of LL GRBs to be $\rho_0^{\rm LL} \sim 200 \rm~Gpc^{-3}~yr^{-1}$ as suggested by Ref.~\cite{Sun:2015bda}. 
Note that the derived $\xi_{\rm CResc}$ is only the lower limit since most of cosmic rays are confined in the sources, and the CR loading factor defined for the total energy of accelerated CRs~\cite{Murase:2005hy} is larger. For a $s_{\rm acc}=2$ spectrum of accelerated cosmic rays, this implies $\xi_{\rm CRacc}\sim100-200$, which is consistent with the previous work~\cite{Murase:2008mr}. 

\subsection{Discussion}
\begin{figure}
\includegraphics[width=\linewidth]{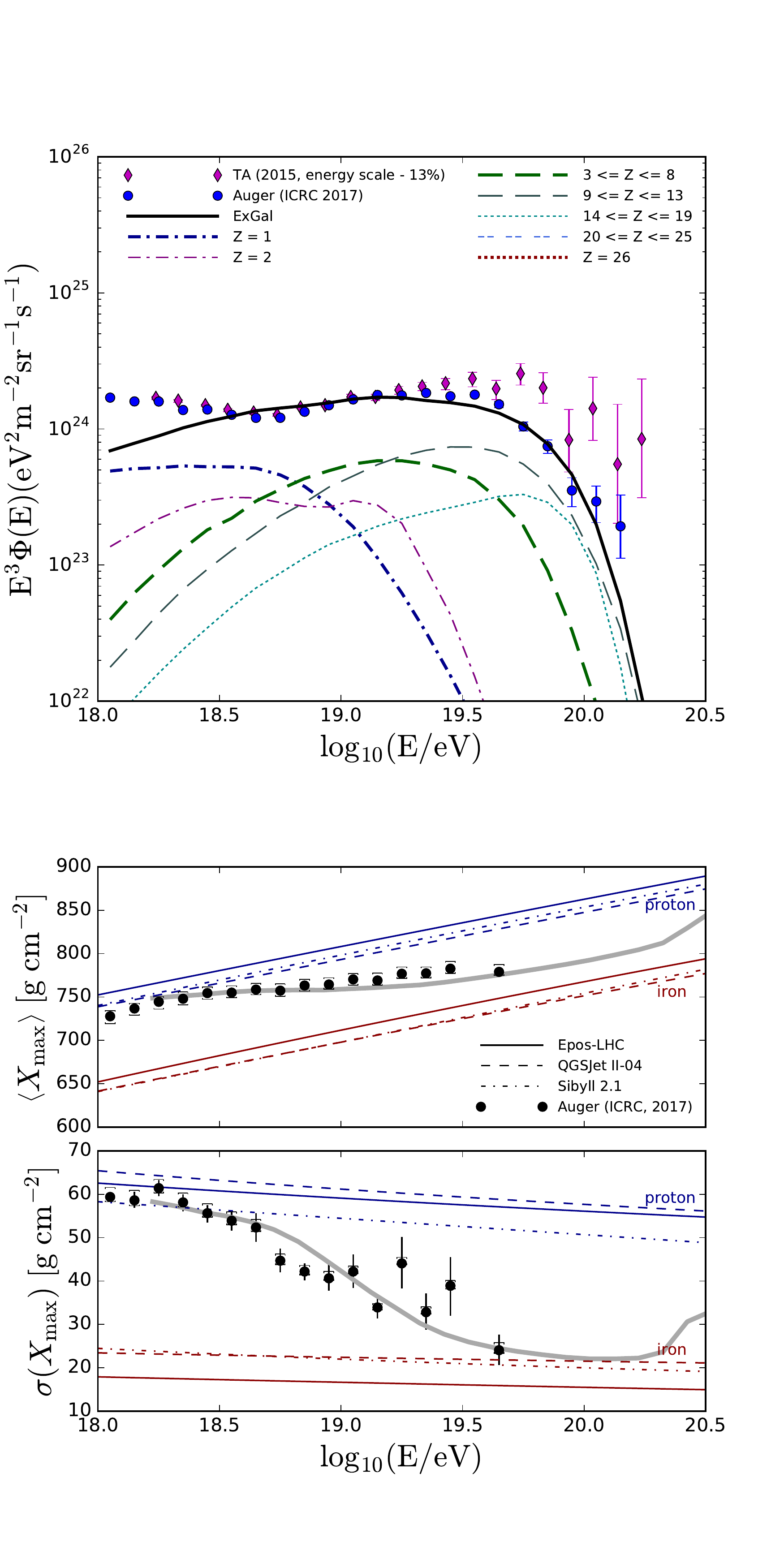}
\caption{The UHECR nuclei spectrum and distribution of $\left\langle X_{\rm max}\right\rangle $ and $\sigma(X_{\rm max})$ calculated from the model Si-R 2 with power-law injection spectrum. The maximum acceleration energy is $Z E_{p, {\rm max}}^\prime = 10^{18.7} Z {L_{\gamma\rm iso, 47}}^{1/2} \rm~eV$, spectral index is $s_{\rm esc} = 0.5$, and $\delta_E = 0.06$. 
\label{fig:16TJ_luminosity_results_powerlaw}}
\end{figure}

To discuss effects of different possibilities of particle escape, we also consider a power-law spectrum of escaping cosmic rays, which is given by
\begin{equation}
\frac{dN_{A^\prime}}{dE^\prime} = f_{A^\prime} N_0 \left( \frac{E^\prime}{Z E_0}\right)^{-s_{\rm esc}} {\rm exp} \left( -\frac{E^\prime}{Z E_{p, {\rm max}}^\prime}\right),
\label{eq:powerlaw}
\end{equation}
where $s_{\rm esc}$ is the spectral index of the escaping cosmic rays and $E_0 = 10^{18}\rm~eV$ is used in this work. 
However, the spectrum of escaping UHECRs depends on details of the escape mechanism. 
To demonstrate that our model can work for power-law spectra, we consider the case of $s_{\rm esc}=0.5$. This could be achieved if the spectrum of accelerated cosmic rays is $s_{\rm acc}=1.5$, which is expected by the first order Fermi acceleration mechanism in the large angle scattering limit (e.g.,~\cite{Aoi:2009ty}) or by magnetic reconnections (e.g.,~\cite{Sironi:2014jfa}).  
See Fig.~\ref{fig:16TJ_luminosity_results_powerlaw}.

\begin{figure}
\includegraphics[width=\linewidth]{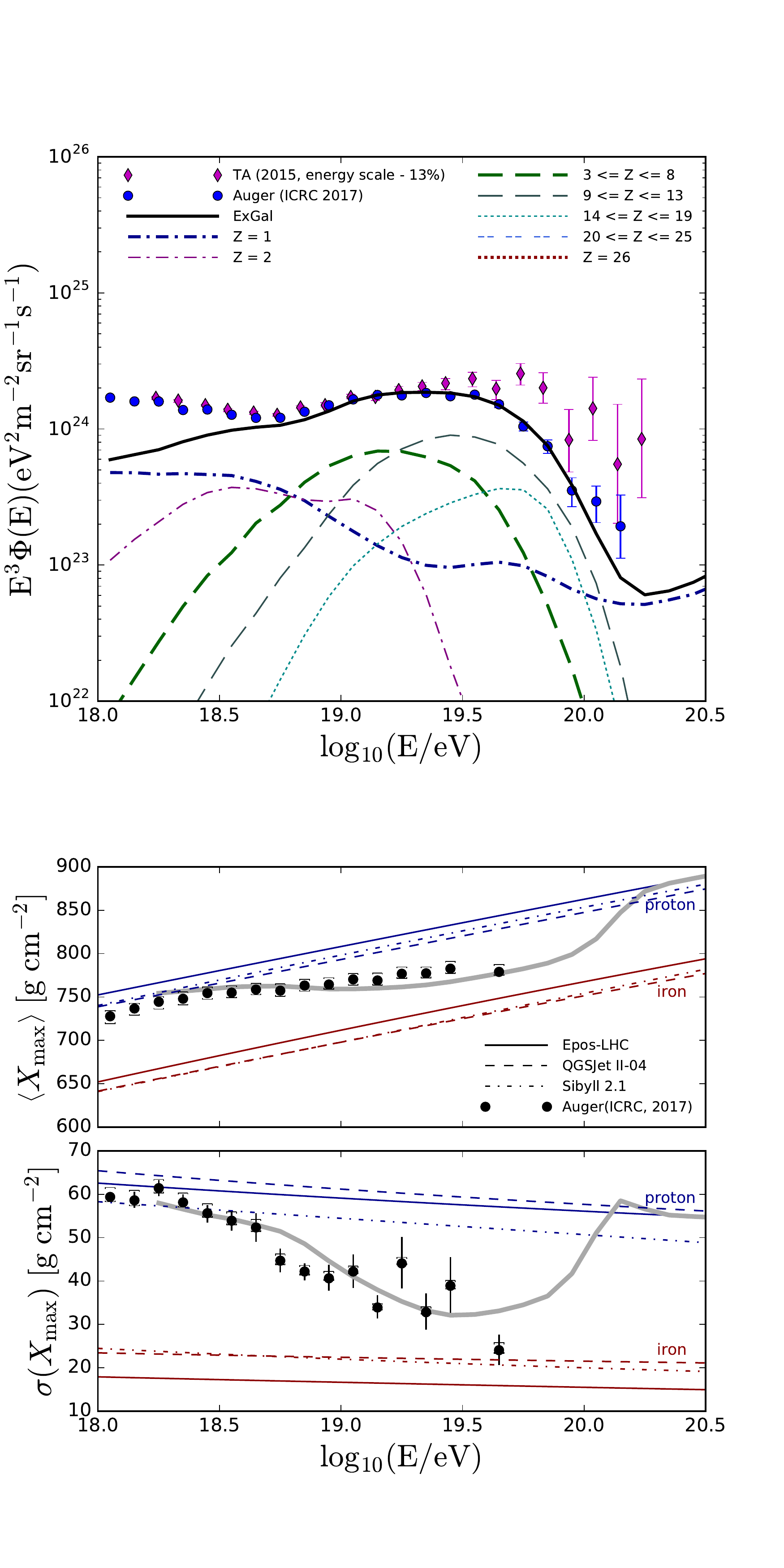}
\caption{Same as Fig.~\ref{fig:16TJ_luminosity_results} but with HL GRB contribution added. The maximum acceleration energy is $Z E_{p, {\rm max}}^\prime = 10^{18.2} Z {L_{\gamma\rm iso, 47}}^{1/2} \rm~eV$ and $\delta_E = 0.14$. 
\label{fig:16TJ_luminosity_results_HLGRB}}
\end{figure}

So far, we only showed the LL GRB contribution. For the purpose of demonstrating the HL GRB contribution, in Fig.~\ref{fig:16TJ_luminosity_results_HLGRB}, we show the case with the HL GRB contribution assuming the proton composition for the HL GRBs and the LL GRB duration 200 times longer than the HL GRB duration. 
The main results are unaffected with the luminosity function used in this work. 
If the composition for the HL GRBs is proton-dominated, the model predicts that the composition changes at the highest energies, $\sim 10^{20.2}\rm~eV$. For a given ratio of LL GRBs to HL GRBs, the HL GRB contribution shown here may be optimistic. In reality, for HL GRBs, the maximum energy of protons would be reduced by energy loss processes. Also, if the jet composition is dominated by nuclei, the maximum energy of UHECR nuclei can be lower due to the photodisintegration process. Note that for HL GRBs nuclei are likely to be destroyed at the jet base in the fireball model, so magnetically dominated jets seem necessary to have a large fraction of nuclei in the jet. The case where UHECR nuclei dominantly come from HL GRBs, was considered by Refs.~\cite{Murase:2008mr, Wang:2007xj} and later by Refs.~\cite{Globus:2014fka,Biehl:2017zlw}. Our LL GRB model can overcome several caveats in the HL GRB model, which are discussed in Section~\ref{sec:three}.  

\section{\label{sec:five} Connection to the IceCube neutrinos}
Murase et al.~\cite{Murase:2006mm} suggested that LL GRBs can be the dominant sources of IceCube's neutrinos (see also \cite{Gupta:2006jm,Liu:2011cua}).
Interestingly, one of the predictions for a low Lorentz factor of $\Gamma=5$ is compatible with the IceCube data above $\sim0.1$~PeV~\cite{Murase:2015xka}, and the medium-energy neutrinos could be explained by their choked jet contribution that can be more abundant~\citep{Murase:2015xka}. The diffuse neutrino flux from high-energy nuclei can be estimated using the simple analytic formula~\cite{Murase:2010gj},
\begin{eqnarray}
E_\nu^2 \Phi_\nu &\approx & \frac{c}{4 \pi H_0} \frac{3}{8} \xi_z f_{\rm sup} {\rm{min}} [1,  f_{p\gamma} (E_A / A) f_{A\gamma} (E_A) \nonumber \\ &+& f_{\rm mes} (E_A) (1 - f_{A\gamma}(E_A))] E_A^2 \frac{dN_A}{dE_A} \rho_0^{\rm LL}  \nonumber \\ &\sim& 2 \times 10^{-8}~{\rm GeV}~{\rm cm^{-2}}~{\rm s}^{-1}~{\rm sr^{-1}} {\rm min}[1, f_{p\gamma}] f_{\rm sup} \nonumber \\ &\times &  \left(\frac{\xi_{\rm CR}/{\mathcal R}}{1}\right)  \left(\frac{\xi_z}{3}\right) \left( \frac{{\mathcal E}_{\rm rad}^{\rm iso}}{10^{50}~\rm erg}\right) \left( \frac{\rho_0^{\rm LL}}{200~\rm Gpc^{-3}~yr^{-1}}\right),\,\,\,\,\,\,
\end{eqnarray}
where the factor $\xi_z$ includes the contribution from high-redshift sources and $f_{\rm sup}$ taking into account the possible effect due to meson and muon cooling. High-energy neutrinos from LL GRBs can be produced by the photomeson production by nuclei (with the effective optical depth $f_{\rm mes}$) and by secondary nucleons (with the effective optical depth $f_{A\gamma}(E_A / A)$) \cite{Murase:2008mr}, and we have used a rough relationship $f_{p\gamma} \sim f_{\rm mes}$ considering that the photomeson production cross section is roughly proportional to $A$. We can see that it is possible for the observed LL GRBs to account for the diffuse neutrino flux observed by IceCube, $\sim3\times10^{-8}\rm~GeV~cm^{-2}~s^{-1}~sr^{-1}$ if $f_{p\gamma} \sim 1$. Note that $f_{p\gamma}\sim1$ implies that nuclei are destroyed and the resulting neutrino flux violates the nucleus-survival bound~\cite{Murase:2010gj}. This implies that the diffuse UHECR flux and neutrino flux can be explained by LL GRBs in the multizone model, where neutrinos come from inner radii such as the photospheric radius and UHECRs originate from outer radii such as the external reverse shock radius~\citep{Bustamante:2014oka,Bustamante:2016wpu}.   
 
We also predict the flux of cosmogenic neutrinos which are produced during the propagation of UHECR nuclei in the intergalactic space due to the interaction with CMB and EBL photons. The cosmogenic neutrino flux is estimated to be $E_{\rm \nu}^2 \Phi(E_{\rm \nu}) \sim 10^{-10}~\rm GeV~cm^{-2} ~s^{-1}~sr^{-1}$. 
Note that this flux of the cosmogenic neutrinos is nearly one or two orders lower than the prediction of the proton dominated scenario (e.g.,~\cite{Beresinsky:1969qj,Yoshida:1993pt,Takami:2007pp,Kotera:2010yn}), so the detection would require ultimate neutrino detectors such as GRAND~\cite{Fang:2017mhl}. On the other hand, the possible contribution from HL GRBs may enhance the cosmogenic neutrino flux if their composition is dominated by protons, in which the neutrino signals may be detected by future neutrino detectors such as ARA~\cite{Allison:2011wk} and ARIANNA~\cite{Barwick:2014pca}.

\section{\label{sec:six} Summary}
In this work, we revisited the UHECR--LL-GRB hypothesis in the case that UHECRs are dominated by intermediate and heavy nuclei. Focusing on LL GRBs, we demonstrated that they can be the promising candidate sources of the UHECR nuclei observed by Auger.

First, we give new predictions based on the intrajet nuclear composition in LL GRBs for different progenitor models under the ``one-time'' injection scenario. In this scenario, the nuclear component in the jet comes from the newly formed accretion disk that is composed of stellar material. The models for the jet nuclear composition can be divided into two groups according to the mass fraction of silicon nuclei, the Si-free models (Si-F 1 (HE 16F), Si-F 2 (16TI)) and the Si-rich models (Si-R 1 (12TJ), Si-R 2 (16TJ) and Si-R 3 (35OC)). See Table.~\ref{tab:table1}. Assuming that the jet base is located at a distance similar to that of the critical radius for disk formation, we confirmed that it is possible for nuclei to survive during the initial loading phase in LL GRBs. Motivated by the connection between LL GRBs and broadline Type Ibc supernovae, we also consider the hypernova ejecta composition which contains a significant fraction of iron nuclei. The detailed physical processes of how nuclei are picked up by the jet are still unclear, which are left as future work.

Then, we discussed the fate of nuclei at the jet base and in the dissipation region (where internal shocks can occur). Our calculations showed that UHECR nuclei easily survive in LL GRBs, while they can be much more easily destroyed into free nucleons in HL GRBs. 
In this work, as a model of the injection spectrum of UHECRs, we adopt the nearly monoenergetic spectrum that is motivated by the escape-limited acceleration model. In this scenario, only cosmic rays around the maximum energy can escape efficiently and the injection spectrum does not have to follow a widely used power-law form. 
Using the procedure similar to that used in Refs.~\cite{Murase:2008mr,Horiuchi:2012by,Zhang:2017hom}, our work confirmed and largely expanded the results obtained by Refs.~\cite{Murase:2008mr,Horiuchi:2012by}, which is also different from the previous works focusing on HL GRBs~\citep{Globus:2014fka,Biehl:2017zlw}.  
We also considered a conventional power-law form, and the Auger spectrum and composition data can be fitted by a hard spectrum with $s_{\rm acc}\lesssim 1.5$. 

We numerically calculate the intergalactic propagation of UHECRs injected by GRB jets. We found that the Si-rich models or hypernova model can nicely account for the Auger data, while it is difficult for the Si-free models to fit the UHECR spectrum. Our results are consistent with the work in Ref.~\cite{Taylor:2015rla, Aab:2016zth}, who suggests that the Auger data can be fitted with sources containing Silicon nuclei. The next-generation detector Auger Prime aims to make more accurate measurements of the UHECR composition~\cite{Aab:2016vlz}, which is essential to test our model.  
We also discussed the effect of the contribution from HL GRBs assuming the proton composition. While the contribution from HL GRBs does not have to affect the results especially for the steeper luminosity function, the composition change is expected at the highest energies. Another prediction of the two population models by HL GRBs and LL GRBs is the transition of the effective number density of the UHECR sources~\cite{Murase:2008sa,Takami:2011nn}. 

Recently, the Pierre Auger Collaboration reported an intriguing indication of the cross correlation between the arrival directions of UHECRs and the catalogues of candidate UHECR sources, where the strongest evidence comes from the correlation with SBGs which has a $4.0\sigma$ statistical significance~\cite{Aab:2018chp}. The host galaxies of HL GRBs are typically low-luminosity blue galaxies, where the blue color implies a high star-formation activity. On the other hand, typical nearby SBGs generally belong to normal or high luminosity star-forming galaxies, although high-redshift SBGs may have lower luminosities.
At present, types of host galaxies of LL GRBs are uncertain since only a handful LL GRBs have been observed~\cite{Liang:2006ci, Virgili:2008gp, Sun:2015bda}. However, since their progenitors are massive stars and their environmental metallicity does not have to be as low as that of HL GRBs, it is possible that LL GRBs and accompanied hypernovae are preferentially hosted by SBGs. 
The energy loss length of silicon nuclei at $\sim10^{20}$~eV is about 100~Mpc. The star-formation rate of nearby 20 SBGs within 50~Mpc is $\sim10$\% of the corresponding total star-formation rate, and the ratio becomes $\sim3$\% for star-forming galaxies within 100~Mpc. 
Thus, assuming that the UHECR energy generation rate simply traces the star-formation rate~\cite{Woosley:2006fn, Savaglio:2008fj}, the contribution from the nearest SBGs is $\sim3-10\%$ of the total UHECR flux. Intriguingly, within an order of the magnitude, this is consistent with the result in Ref.~\cite{Aab:2018chp}, where the best-fit result suggests that $\sim10\%$ of the observed UHECRs come from the nearby catalogued 23 SBGs. In this sense, our model gives one of the viable explanations why the observed UHECRs are correlated with the SBGs.


\medskip
\begin{acknowledgments}
We thank Denise Boncioli, Karl-Heinz Kampert, Michael Unger, and Alan Watson for useful comments and discussions. 
The work of K.M. is supported by Alfred P. Sloan Foundation and the U.S. National Science Foundation (NSF) under grants NSF Grant No. PHY-1620777. S.H.~is supported by the U.S.~Department of Energy under award number de-sc0018327. P.M. is supported by NASA NNX13AH50G (P.M.). S.S.K. acknowledges IGC fellowship at Pennsylvania State University. B.T.Z. is supported by the China Scholarship Council (CSC) to conduct research at Penn State University and the High-performance Computing Platform of Peking University.
\end{acknowledgments}

\bibliography{bzhang}

\end{document}